\documentclass[a4paper,11pt]{article}
\usepackage{pos}
\usepackage{xspace}
\usepackage{journalnames}
\usepackage{bibspacing}
\newcommand{\pdx}[1]{_{\rm #1}}

\def\pdot {\dot P}

\def\edot {\dot E}

\def\sps{s\,s$^{-1}$\xspace}

\def\flux{erg\,s$^{-1}$\,cm$^{-2}$\xspace}
\def\lum{erg\,s$^{-1}$\xspace}

\def\deg{^\circ}

\def\xds{XDINSs\xspace}
\def\xd{XDINS\xspace}

\newcommand{\psrj}{PSR J0726$-$2612\xspace}
\newcommand{\rxj}{RX J1856.5$-$3754\xspace}

\def\xmm{{\em XMM-Newton}\xspace}

\def\chandra{{\em Chandra}\xspace}
\def\rosat{{\em ROSAT}\xspace}
\def\erosita{{\em eROSITA}\xspace}
\def\nicer{{\em NICER}\xspace}
\def\ixpe{{\em IXPE}\xspace}

\usepackage{color}

\title{X-ray observations of Isolated Neutron Stars}
 \ShortTitle{X-ray observations of INS}

\author*[a]{Michela Rigoselli}

\affiliation[a]{INAF, IASF Milano,\\
 Via Alfonso Corti 12, I-20133 Milano, Italy}

\emailAdd{michela.rigoselli@inaf.it}

\abstract{
Pulsars are rapidly spinning neutron stars, that radiate at the expense of their strong magnetic field and their high surface temperature.
Five decades of multi-wavelength observations showed a large variety of 
physical parameters, such as the spin period, the magnetic field and the age, and 
of observational properties, especially in the radio and X-ray band.
Isolated neutron stars have been classified according to the presence of thermal or non-thermal emission, and whether they show a constant flux, rapid flares and bursts or long-standing outbursts.
One of the current challenges in the study of such objects is to explain these different manifestations in the context of a unified evolutionary picture.
On the other hand, recent findings show that the classes of isolated neutron stars are more connected than previously thought, and that non only magnetars hold a complex magnetic field topology in the crust and above the surface.
}

\FullConference{%
Multifrequency Behaviour of High Energy Cosmic Sources - XIV\\
12--17 June 2023\\
Palermo, Italy\\
}

\begin{document}

\maketitle

\section{The Isolated Neutron Stars zoology}
Neutron stars have been detected for the first time as radio pulsating sources in 1967 \citep{hew68}, and in the subsequent fifty years more than 3300 have been registered\footnote{\href{https://www.atnf.csiro.au/research/pulsar/psrcat/}{https://www.atnf.csiro.au/research/pulsar/psrcat/}.} \citep{2005AJ....129.1993M}. They have mainly been discovered thanks to the detection of their pulsed non-thermal emission, at wavelengths spanning from radio to $\gamma$-rays, and they can be isolated stars or members of a binary system.

The energy that sustains pulsar emission is supplied by their fast rotation  (Section~\ref{sec:rpp}), via the braking operated by their intense magnetic field, that is assumed to be dipolar.
Under these very simple assumptions, three characteristic quantities can be inferred from the period $P$ and its derivative $\pdot$: the so-called spin-down luminosity
\begin{equation}
    \dot{E}_{\rm rot} = 4 \pi^2 I_{\rm NS} \dot{P} P^{-3},
    \label{eq:erot}
\end{equation}
where $I_{\rm NS} \approx 10^{45}$ g cm$^2$ is the moment of inertia, the dipolar magnetic field on the surface
\begin{equation}
    B_{\rm dip} \approx 3.2 \times 10^{19}\, (P \dot{P})^{1/2}~\rm G,
    \label{eq:bdip}
\end{equation}
and the characteristic age
\begin{equation}
    \tau_c = \frac{P}{2\dot{P}},
    \label{eq:tauc}
\end{equation}
that takes this form if the current period is way larger than the initial period $P_0$.

With the advent of X-ray satellites and all-sky surveys, several thermally emitting isolated neutron stars (INSs) have been discovered at the center of supernova remnants (SNRs) (Section~\ref{sec:cco}), or as serendipitous sources with a very soft X-ray stable spectrum (Section~\ref{sec:xd}), or as extremely flaring sources (Section~\ref{sec:mag}).

\begin{figure*}[!ht]
\centering
\includegraphics[width=1\textwidth]{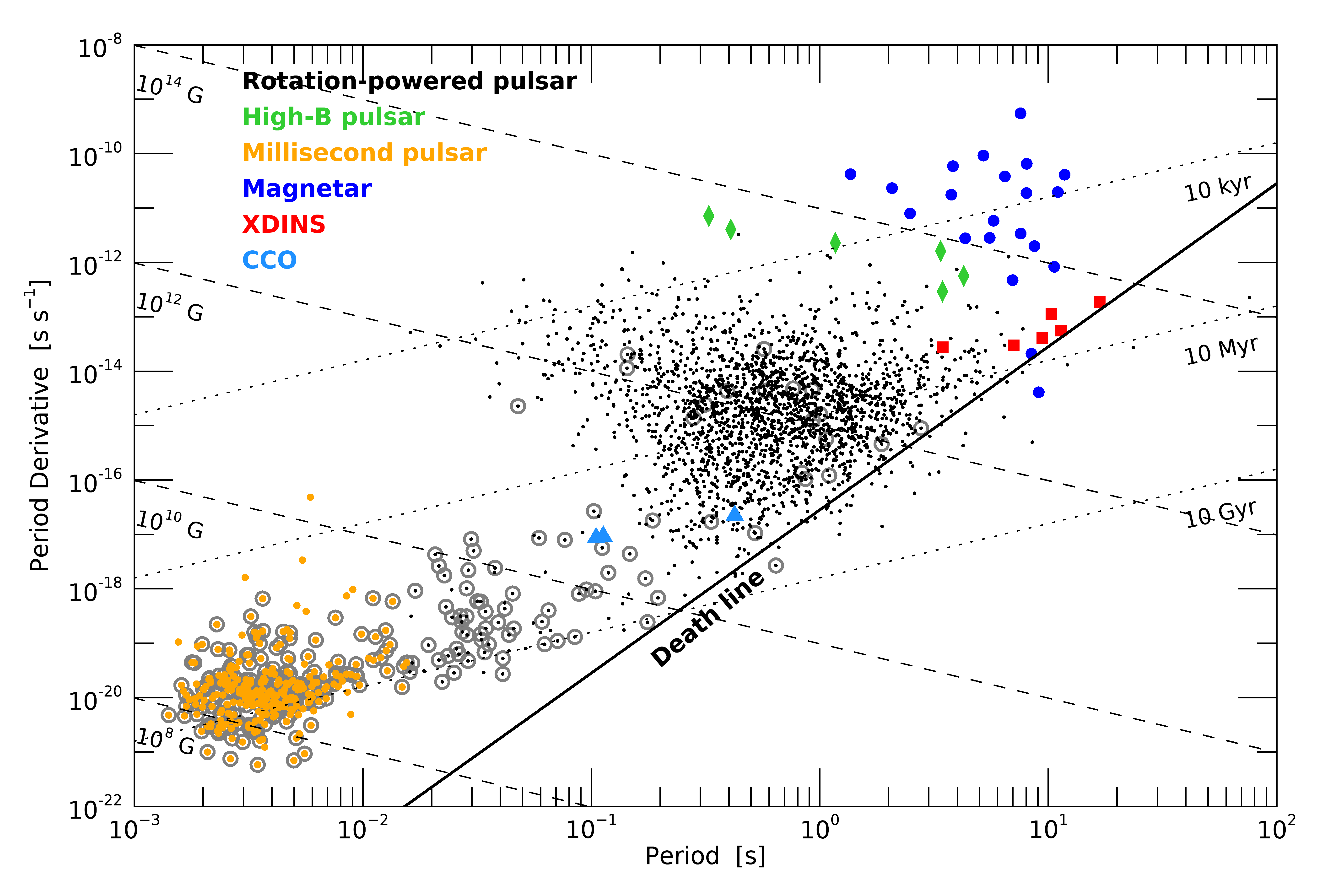}
\caption{\small $P-\pdot$ diagram of pulsars. The bulk of the pulsar population is made up of the rotation-powered pulsars (black dots), sub-divided into high-B pulsars (green diamonds) and millisecond pulsars (orange dots). A significant fraction of the pulsars ($10\%$) are in a binary system (grey circles). The other classes of INSs plotted here the magnetars (blue dots), the \xds (red squares) and the CCOs (light blue triangles). Lines of equal characteristic age (dotted, $10^{4} - 10^{10}$ yr) and equal dipole magnetic field (dashed, $10^{8} - 10^{14}$ G) are indicated. The radio pulsar death line $B/P^2 = 1.7 \times 10^{11}$ G s$^{-2}$ \citep{bha92} is also shown. The data are taken from the ATNF Pulsar Catalogue \citep{2005AJ....129.1993M}.} 
\label{fig:PPdot}
\end{figure*}

$P$ and $\pdot$ play a fundamental role in characterizing the pulsar properties, and the neutron star population is usually represented in the $P-\pdot$ diagram (shown in Figure~\ref{fig:PPdot}), as the ordinary stars are represented in the Hertzsprung-Russell diagram; the different classes of INSs are placed on different zones of the diagram.

In the following, I will review the main X-ray properties of INSs and the links that are emerging among the different classes. On this topic, see also \citep{2013FrPhy...8..679H,2014AN....335..262I,2018IAUS..337....3K,2023Univ....9..273P}.

\subsection{Rotation-powered pulsars}\label{sec:rpp}

The rotation-powered pulsars (RPPs) are the bulk of INSs and are detected because of their pulsed emission.
Nowadays more than 3000 RPPs are known, and they have been detected from the radio band to the very-high energy $\gamma$-rays (see e.g.\ the recent result on Vela pulsating emission at 20 TeV \citep{2023NatAs.tmp..208H}). They also show optical/UV/X-ray thermal emission from the cooling surface and from hot spots heated by returning currents \citep[and references therein]{pot20}.

RPPs fill the central region of the $P-\pdot$ diagram (Figure~\ref{fig:PPdot}, black dots); a newborn pulsar appears in the top-left corner and, according to the magnetic-dipole braking model, it evolves along the $B\pdx{dip}$ constant lines (dashed), crossing the $\tau_c$ constant lines (dotted). The bottom-right corner of the plot is empty because when pulsars cross the so-called ``death line'' \citep{bha92}, they are too old and too slow 
to maintain the required potential difference for pair production in the vacuum gap \citep{bar98,bar01}.


Actually, there is a population of RPPs that have $\tau_c \gtrsim 1$ Gyr that are still detected: they are known as millisecond pulsars (MSPs) because of their short and stable period ($P \lesssim 16$ ms \citep{2023RNAAS...7..213H}, $\pdot \lesssim 10^{-18}$ s s$^{-1}$, see Figure~\ref{fig:PPdot}, orange dots). 
They are old pulsars that have been spun up through the accretion of matter from the binary companion at some time in the past, as part of a binary system. The majority of them still have a companion star and they emit powerful non-thermal emission in the radio and X-ray band, plus thermal X-rays from heated polar caps (see e.g.\ \citep{man17} for a review).


Among the young RPPs, there is a sub-sample of objects having magnetic fields larger than
\begin{equation}
    B\pdx{QED} = \frac{m_e^2c^3}{e \hbar} \approx 4.4\times10^{13} \mathrm{~G}
    \label{eq:bqed}
\end{equation}
(see Figure~\ref{fig:PPdot}, green diamonds).
They are usually referred to as high-B pulsars (see the review \citep{ng11}), and they do not constitute a separated class of INSs, but they act as ordinary pulsars most of the time, and suddenly show bursting phenomena as the magnetars (see Section~\ref{sec:mag}) or share the spectral properties of the XDINS (see Section~\ref{sec:psrj}).

\subsection{Magnetars}\label{sec:mag}


Magnetars\footnote{A catalog of magnetars can be found at \href{http://www.physics.mcgill.ca/\~pulsar/magnetar/main.html}{http://www.physics.mcgill.ca/$\sim$pulsar/magnetar/main.html}, see also \citep{2014ApJS..212....6O}.}  are INSs that have been discovered thanks to their repeated bursting activity\footnote{See also \href{http://magnetars.ice.csic.es/\#/outbursts}{http://magnetars.ice.csic.es/\#/outbursts} for a complete catalogue of outbursting magnetars.} \citep{2018MNRAS.474..961C} and pulsating emission ($P\sim2-12$ s, $\pdot \sim 10^{-9}-10^{-12}$ \sps) in the X-ray band. It was soon clear that they could not be powered by rotation, as their X-ray luminosity exceeds $\edot_{\rm rot}$ by some orders of magnitude, but by their very strong magnetic field \citep{DT92,DT95,DT96} ($B\pdx{dip} \sim 10^{14}-10^{15}$ G, see Figure~\ref{fig:PPdot}, blue dots). Recent reviews on magnetars are \citep{2017ARA&A..55..261K,2021ASSL..461...97E}, see also Andrei Beloborodov's contribution.

The persistent X-ray emission of magnetars is characterized by soft thermal spectra ($kT \sim\!0.2-0.6$ keV) plus a power-law component in the hard X-ray band with photon index $\Gamma=1-4$, commonly interpreted as resonant Compton scattering of thermal photons by charged particles flowing along the closed field lines of the twisted magnetosphere \citep{2002ApJ...574..332T}. The continuum is often absorbed by absorption features \citep[e.g.]{2003ApJ...586L..65R,2014AN....335..274T} that, if interpreted as proton cyclotron features provide a magnetic fields
\begin{equation}
    B\pdx{cyc,p} = E\pdx{abs} \frac{m_p c} {\hbar e}(1 + z)^{-1} \approx  \frac{E\pdx{abs}}{1~ \mathrm{keV}} \, 1.3 \times 10^{14}~\mathrm{G},
    \label{eq:Bcycp}
\end{equation}
where $z\approx0.2$ is the gravitational redshift.

Recently, polarization patterns from the three brightest magnetars in quiescence were detected between $2-8$ keV by the \ixpe satellite \citep{2022Sci...378..646T,2023ApJ...944L..27Z,2023ApJ...954...88T}, see also Roberto Taverna's contribution.
In the radio band only few magnetars are detected, and they are characterized by large variability both in flux and pulse profile shape on timescale of days, by a very flat spectrum ($\alpha > -0.5$) and high polarization \citep{cam07,cam08}.

Recent discoveries challenged the long standing belief that magnetars must posses supercritical magnetic fields (Eq.\ \ref{eq:bqed}): 
there are radio pulsars that sometimes show bursting activity (PSR J1846$-$0258 \citep{2008Sci...319.1802G}, PSR J1119$-$6127 \citep{2016ApJ...829L..21A})
and there are neutron stars acting like magnetars but with a dipolar magnetic field lower than $4\times10^{13}$ G (SGR J0418$+$5729 \citep{2010MNRAS.405.1787E}, Swift J1822$-$1606 \citep{2012ApJ...754...27R} and 3XMM J1852$+$0033 \citep{2014ApJ...781L..17R}).
This topic was considered in more details in \citep{2017hsn..book.1401B}.
Detailed calculations show that a local magnetic field $>\!10^{14}$ G should be necessary to break the crust, in order to produce bursts and outbursts \citep{2015MNRAS.446.1121G,2015MNRAS.449.2047L}; these values can be provided by the internal toroidal components and higher-order multipoles \citep[e.g.][]{2011ApJ...740..105T}. For this reason, objects with similar dipolar magnetic field strength as inferred from their timing parameters can display very different behaviors.

\subsection{Central Compact Object}\label{sec:cco}
The central compact objects (CCOs) form a class of INSs that are found in the center of young ($0.3-7$ kyr) SNRs, that emit only in the X-rays and have no counterparts at any wavelength \citep{mig19}. A review on these objects can be found in \citep{2017JPhCS.932a2006D}.

At present, the class counts a dozen objects, three of which are pulsators \citep{zav00,got05,got09}. The CCOs are detected in the soft X-ray range, and their spectra are exclusively thermal, with high temperatures ($0.2-0.5$ keV) and very small emitting radii (ranging from 0.1 to a few km). Two sources show absorption features at $0.7-0.8$ keV \citep{big03,got10}, that have been interpreted as electron cyclotron lines due to a magnetic field of the order of
\begin{equation}
    B\pdx{cyc,e} = E\pdx{abs} \frac{m_e c} {\hbar e}(1 + z)^{-1} \approx  \frac{E\pdx{abs}}{0.1~ \mathrm{keV}} \, 7.2 \times 10^{9}~\mathrm{G}.
    \label{eq:Bcyce}
\end{equation}

The three pulsating CCOs (see Figure~\ref{fig:PPdot}, light blue triangles) have periods of $0.1-0.4$ s and spin derivatives of about $10^{-17}$ s~s$^{-1}$ \citep{hal10,got13}, from which weak dipole magnetic fields ($B\pdx{dip} \sim 10^{10}$ G) and high characteristic ages ($\tau_c \sim 10^{8}$ yr) are derived. This is at variance with the SNR associations, and the reason could be that the approximation of Eq.\ \ref{eq:tauc} is no longer valid because these sources have $P \approx P_0$.

Also the picture of CCOs as weakly magnetized INSs has issues: the high pulsed fraction (up to 64\% \citep{hal10}) and the steep surface temperature distribution cannot be easily explained without invoking magnetic fields of $10^{14}-10^{15}$ G \citep{sha12} (see also Section~\ref{sec:thx}). To explain this tension, the ``buried B scenario'' was invoked: the actual strong field is screened by the matter which fallback onto the star after the supernova explosion \citep[][and references therein]{2020MNRAS.495.1692G}.
One remarkable CCO in SNR RCW 103 shows a 6.7-hr X-ray periodicity of yet unknown origin as well as distinctly magnetar-like
behavior \citep{2016ApJ...828L..13R,2016MNRAS.463.2394D}. Such long spin periods in young non-accreting objects can be explained in a model where a strong magnetic field of the star interacts with a fallback disc (see Section \ref{sec:longp}).

\subsection{X-ray-dim Isolated Neutron Stars}\label{sec:xd}
Among INSs, the so-called X-ray-dim isolated neutron stars (\xds) represent a peculiar class of nearby sources, characterized by their thermal emission in the X-ray band, with faint optical/UV counterparts \citep{2011ApJ...736..117K} and no confirmed detection of radio pulsations \citep{2009ApJ...702..692K}. Reviews on \xds can be found, e.g., in \citep{2007Ap&SS.308..191V,2009ASSL..357..141T}.

They are seven INSs discovered in the nineties by the \rosat satellite and soon gained the nickname of ``Magnificent Seven''.
The \xds have spin periods in the range $P\sim 3-17$ s and period derivatives of a few $10^{-14}$ s\,s$^{-1}$ (see Figure~\ref{fig:PPdot}, red squares), which result in characteristic ages of $\tau_c \sim 1-4$ Myr and magnetic fields of the order of a few $10^{13}$ G.

Their very soft ($kT\lesssim 0.1$ keV) X-ray spectra are well reproduced by a simple blackbody with little interstellar absorption, with the additional presence of broad absorption lines at $0.2-0.4$ keV in most sources \citep{2003A&A...403L..19H, 2004ApJ...608..432V,2004A&A...424..635H,2005ApJ...627..397Z}, and narrow, phase-variable ones in few cases \citep{2015ApJ...807L..20B,2017MNRAS.468.2975B}.
If these lines are interpreted as proton cyclotron features or atomic transitions \citep[e.g.]{2008AIPC..968..129K}, and if the variable X-ray radiation is modeled taking into account anisotropic thermal conductivity \citep{2020MNRAS.497.2883K}, the magnetic fields estimated are of the same order as those derived from the spin-down rate assuming magnetic dipole braking.

\subsection{The magneto-thermal unification}
The variety of the observational appearance of young INSs must find an explanation, given that they cannot be different classes: the sum of their birthrates exceeds the rate of Galactic core-collapse supernovae, which is about $1/60$ yr$^{-1}$ \citep{2021A&A...646A.117R,2008AIPC..968..129K}.

The idea is to find a combination of initial distributions and evolutionary laws that allows to unite all known types of sources in one framework; this must also include transitions between different types of activity and appearance of hybrid behavior.
Nowadays it is believed that such scenario must include magnetic field evolution, namely its decay and a possible re-emergence on a time scale $\sim\! 10^4 -10^5$ yr, and a significant contribution of non-dipolar fields. 

In a few words, the magnetic field evolves in the solid crust decaying due to the Ohmic dissipation and the Hall drift \citep{hol02,hol04,cum04}. The timescales of these two processes are:
\begin{equation}
    \tau_{\rm Ohm} = \frac{4\pi \sigma L^2}{c^2} \approx 4.4\times 10^6 \left( \frac{\sigma}{10^{24}\rm{~s^{-1}}} \right) \left( \frac{L}{1\rm{~km}} \right)^2 \rm{~yr}
    \label{eq:ohm}
\end{equation}
\begin{equation}
    \tau_{\rm Hall} = \frac{4\pi n_e e L^2}{c B} \approx 6.4\times 10^4 \left( \frac{n}{10^{34}\rm{~cm^{-3}}} \right) \left( \frac{B}{10^{13}\rm{~G}} \right)^{-1} \left( \frac{L}{1\rm{~km}} \right)^2 \rm{~yr}
    \label{eq:hall}
\end{equation}
where $\sigma$ is conductivity and $L$ is the characteristic length of magnetic field variations. 

On the other hand, the surface temperature lowers because of the neutron star cooling.
Given that these two quantities are related to each other and evolve together \citep{gon10}, they have to be studied in a comprehensive auto-consistent model, called ``magneto-thermal evolution'' \citep[and reference therein]{2013MNRAS.434..123V}: temperature affects crustal electrical resistivity, which in turn affects magnetic field evolution, while the decay of the field can produce heat that then affects the temperature evolution.

There is quite general agreement that the magnetic field of relatively young neutron stars decays on timescales of $10^4-10^6$ years, depending on the conductivity, thickness of the crust, and strength and structure of the initial field \citep{2014MNRAS.444.1066I}. 
On the other hand, population synthesis studies \citep{1992A&A...254..198B} suggest that old pulsars show no significant magnetic field decay over their life time (timescales larger than $10^9-10^{10}$ years \citep{2001A&A...376..543T}). 
This issue remains somewhat open, but can be tentatively solved by the assumption that the magnetic field is maintained by two current systems: long living currents in the superconducting core support the large scale dipolar field and are responsible for the spin down of old pulsars, while currents in the crust support the short living part of the field \citep{2007A&A...470..303P}.

Recent X-ray observations provided new evidences of link between the classes of INSs, and challenge the current theoretical models.

\section{Links between \xd and RPP}\label{sec:n2}
The link between \xd and RPP has been reinforced when a radio high-B pulsar was observed sharing timing and spectral properties with the \xds (Section~\ref{sec:psrj}), and when weak non-thermal emission was detected in the cumulative X-ray spectra of two \xds (Section~\ref{sec:rxj}).

\subsection{The case of \psrj}\label{sec:psrj}

\psrj is slowly rotating ($P = 3.4$ s), highly magnetized ($B\pdx{dip} = 3\times10^{13}$ G)
radio pulsar: its timing parameters are in the range of those of the \xds, but it does show radio pulsations. The similarity with the \xds was reinforced by X-ray observations with the \chandra  satellite \citep{2011ApJ...743..183S}, that revealed a soft thermal spectrum and pulsations with a sinusoidal, double-peaked profile with a pulsed fraction of $30 \%$. 

Rigoselli et al.\ 2019 \citep{2019A&A...627A..69R} analyzed deep \xmm data of \psrj and reported that the soft X-ray spectrum can be well fit by two blackbodies (temperatures $kT_1 = 74_{-11}^{+6}$ eV and $kT_2 = 140_{-20}^{+40}$ eV, radii $R_1=10.4_{-2.8}^{+10.8}$ $d\pdx{kpc}$ km and $R_2=0.5_{-0.3}^{+0.9}$ $d\pdx{kpc}$ km) plus a Gaussian line in absorption with the line placed at $E=0.39_{-0.03}^{+0.02}$ keV and with a broadening of $\sigma=0.08_{-0.02}^{+0.03}$ keV. The inferred magnetic field is $B \simeq 5\times 10^{13}$ G (see Eq.\ \ref{eq:Bcycp}) is in good agreement with the dipole magnetic field.

Assuming a distance of 1 kpc, the luminosity of \psrj is $L = (4.0_{-1.0}^{+4.4}) \times 10^{32}$ \lum. This is greater than its spin-down luminosity, as for the \xds, but is in reasonable agreement with the expected thermal luminosity of a $\sim \! 200$ kyr-old pulsar.

Rigoselli et al.\ 2019 \citep{2019A&A...627A..69R} also tried to model the double-peaked pulse profile, but no simple model based on blackbody emission could reproduce the observed pulse profile.
Whatever the mechanism responsible for the surface emission, the presence of a strong magnetic field results in some degree of anisotropy in the emitted radiation. In the case of a magnetized atmosphere, more complicated energy-dependent beaming patterns are produced. The best match with the data was obtained assuming emission from two antipodal hot spots with an effective temperature of 0.5 MK, and $(\Omega - \mu) \approx (\Omega - \mathrm{LOS})$, where $\Omega$ and $\mu$ are the rotation and the magnetic axes, respectively, and the LOS is the direction of the line of sight.

\begin{figure*}[!ht]
\centering
\includegraphics[width=0.45\textwidth]{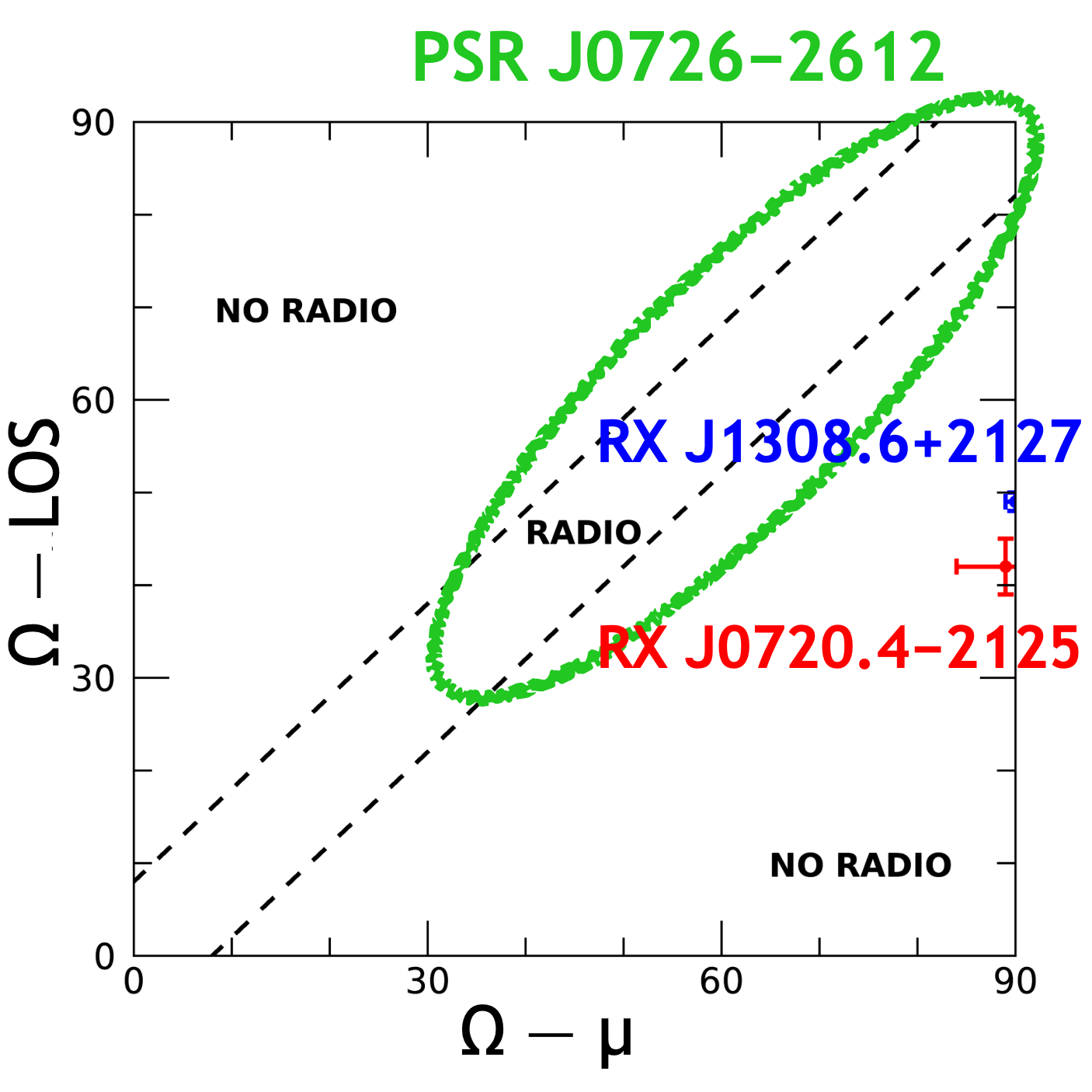} ~\qquad
\includegraphics[width=0.45\textwidth]{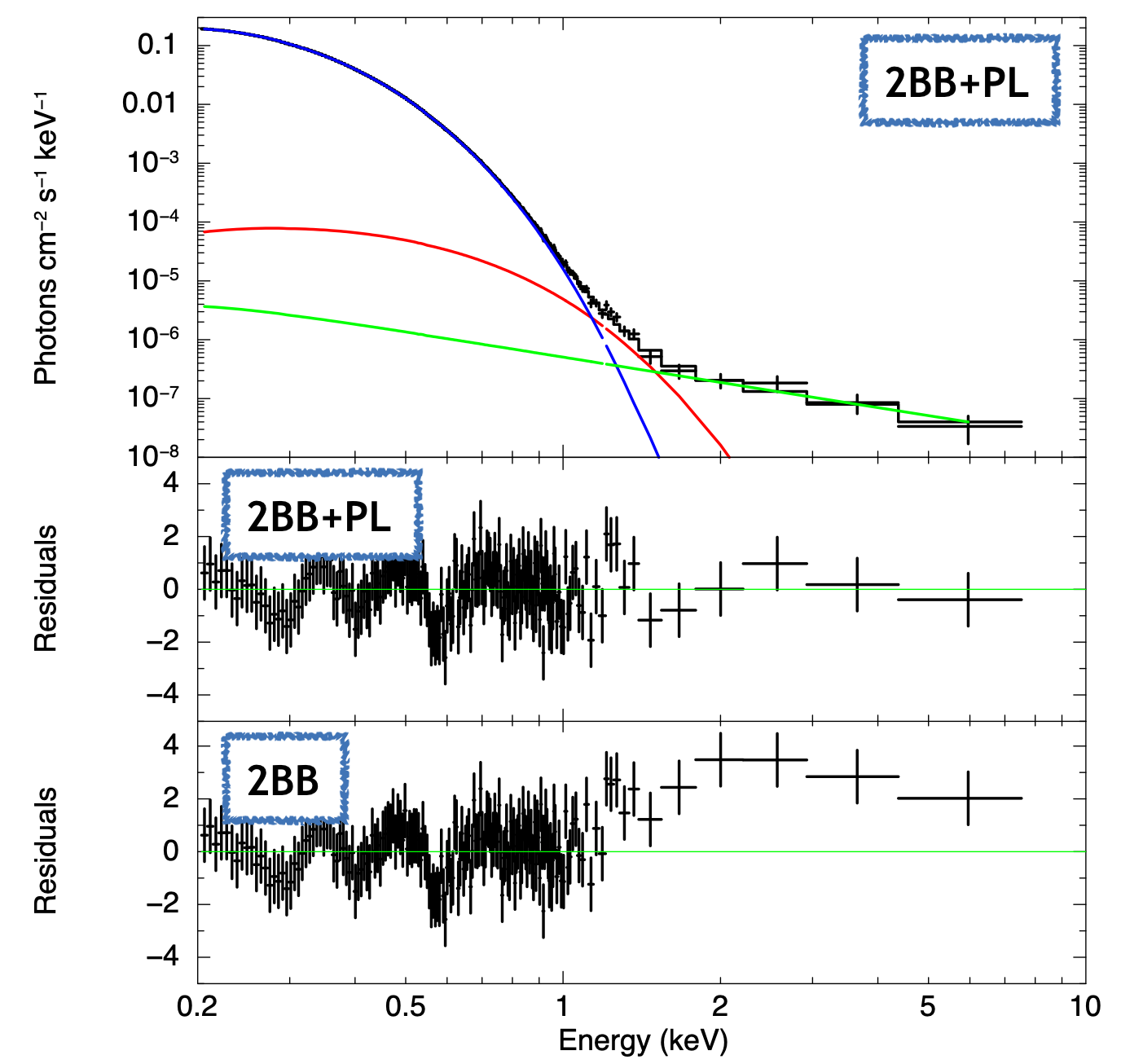}
\caption{\small Left panel: Visibility of a radio beam with aperture of $\sim\!8\deg$ as a function of the angles between $\Omega-\mu$, and $\Omega-$\,LOS. The estimated positions for \psrj, RX J0720.4$-$3125 and RX J1308.6$+$2127 are shown (see Section~\ref{sec:psrj}). Adapted from \citep{2019A&A...627A..69R}. Right panel: Total X-ray spectrum of \rxj obtained with \xmm data (see Section~\ref{sec:rxj}). The best-fitting model with two blackbody (blue and red) and a power-law (green) components is superimposed in the top panel. The two bottom panels show the residuals for the two-blackbody plus power law fit (2BB+PL) itself and for the two blackbody (2BB) fit, which yield unsatisfactory residuals at $E > 1$ keV. On the contrary, the residuals below 0.6 keV are at a level $<\!2\%$, well below the instrument effective area uncertainties. Adapted from \citep{2022MNRAS.516.4932D}.} 
\label{fig:psrxd}
\end{figure*}

While most of the \xds have single-peaked pulse profiles, two of them (RX J0720.4$-$3125 \citep{2011A&A...534A..74H} and RX J1308.6$+$2127 \citep{2017A&A...601A.108H}) show double-peaked profiles similar to \psrj, although with smaller pulsed fractions ($18 \%$ and $11 \%$, respectively). The inferred geometries for these two objects are $(\Omega - \mu) \approx 90\deg$ and $(\Omega - \mathrm{LOS})\approx 45\deg$; with the usual assumption that the radio beam coincides with, or is close to, the magnetic dipole axis, such a large impact parameter can naturally account for the fact that their radio emission is not visible from the Earth. Contrary to the two \xds, \psrj should lie inside this region and this might explain why it is detected in the radio band, while the two \xds are not (see Figure~\ref{fig:psrxd}, left panel).

\subsection{The case of \rxj}\label{sec:rxj}

The soft X-ray pulsar \rxj  is the brightest ($F_X\simeq1.5 \times 10^{-11}$ \flux \citep{1996Natur.379..233W}) and closest ($d=123^{+11}_{-15}$ pc, \citep{2010ApJ...724..669W}) member of the \xd class.
Its pulsation, at $P=7.05$ s  \citep{2007ApJ...657L.101T} and $\pdot=3\times10^{-14}$ \sps \citep{2008ApJ...673L.163V}, was detected despite a very small pulsed fraction $\rm{PF}\simeq1.2\%$. The timing parameters yield $\edot\pdx{rot}=3.4\times10^{30}$ \lum, $\tau_c = 3.7 \times 10^6$ yr and $B\pdx{dip} = 1.5\times10^{13}$ G.
Its X-ray spectrum resembles a pure blackbody emission with temperature $kT \simeq 60$ eV
, even though the emission in the optical band ($V \sim 25.7$ \citep{1997A&A...318L..43N}) lies above the extrapolation of a single-temperature X-ray blackbody at low energies. 

The brightness, simple spectrum, and steadiness of its emission make \rxj an ideal target for the calibration of \xmm, that observed it about every six months since 2002. Using all the 2002--2022 data from the \xmm EPIC-pn camera, De Grandis et al.\ 2022 \citep{2022MNRAS.516.4932D} obtained a $0.3-7.5$ keV spectrum having 1.43 Ms of net exposure time, that showed a hard excess with respect to the pure blackbody emission \citep[see also][]{2017PASJ...69...50Y,2020ApJ...904...42D}. They report a best fit made up of the sum of two blackbodies (temperatures $kT_1 = 61.9 \pm 0.1$ eV and $kT_2 = 138 \pm 13$ eV, radii $R_1 = 4.92_{-0.06}^{+0.04}$ km and $R_2  = 31_{-16}^{+8}$ m) and a power law (photon index $\Gamma = 1.4_{-0.4}^{+0.5}$, flux in the $2-8$ keV band of $(2.5_{-0.6}^{+0.7})\times 10^{-15}$ \flux (see Figure~\ref{fig:psrxd}, right panel).

The luminosity of the non-thermal component corresponds to $10^{-3}$ times the spin-down power $\edot\pdx{rot}$. This value is consistent with what is observed in rotation powered X-ray pulsars with higher $\edot\pdx{rot}$ \citep[e.g.][]{2002A&A...387..993P}, so that a magnetospheric origin for this component appears as the most natural option. Hints for pulsations above 2 keV were reported, and this  would reinforce the magnetospheric origin and the link to the RPPs.

De Grandis et al.\ 2022 \citep{2022MNRAS.516.4932D} analyzed also 20 years of \xmm data of RX J0420.0$-$5022, the \xd having the second highest $\edot_{\rm rot}$ of the class.
They found a hard excess that can be fit either with a second blackbody of a power law, or their sum. Also in this case, the putative non-thermal component would have an efficiency $L\pdx{PL}/\edot\pdx{rot} \sim 10^{-3}$.
Can we expect that the other five \xds have a non-thermal component too? If we assume a similar efficiency and we consider the $\edot\pdx{rot}$ of the remaining stars, the expected X-ray flux should be $F\pdx{PL} \lesssim 10^{-16}$ \flux. This level of flux is beyond the sensitivity of the current facilities, but maybe \emph{Athena} could achieve it\footnote{\href{https://www.cosmos.esa.int/web/athena/study-documents}{https://www.cosmos.esa.int/web/athena/study-documents}.}.

\section{New (?) flavors of INSs}
Thanks to X-ray (Section \ref{sec:newX}) and radio (Section \ref{sec:longp}) current facilities, new flavors of INS have been discovered. The effort is now to make them fit into the classical magneto-thermal evolutionary model.

\subsection{Thermally emitting INSs}\label{sec:newX}
Calvera (1RXS J141256.0$+$792204) is an enigmatic X-ray pulsar with properties that do not fit easily with those of the known classes of INSs. 
It was discovered in the \rosat all sky survey as a soft X-ray source  with high X-to-optical flux ratio, qualifying it as an INS candidate \citep{2008ApJ...672.1137R}.   Its spectral properties (thermal X-ray only, absence of radio or $\gamma$-ray counterpart) resemble those of \xds (Section~\ref{sec:xd}) and CCOs (Section~\ref{sec:cco}), and for these reasons it was initially considered as a possible  new member of the ``Magnificent Seven'' class and nicknamed ``Calvera''. However, it was later discovered that Calvera has a spin period of 59 ms \citep{2011MNRAS.410.2428Z} and is spinning down at a rate   $\pdot=3.2\times10^{-15}$ \sps  \citep{2013ApJ...778..120H}. These timing parameters give a characteristic age $\tau_c = 2.9\times10^5$~yr and  a dipole magnetic field $B_{\rm dip}=4.4\times10^{11}$ G, similarly to middle-aged RPPs.

A magnetized hydrogen atmosphere model, covering the entire star surface and having an anisotropic temperature map, provides a good description of the phase-resolved spectra and energy-dependent pulsed fraction \citep{2021ApJ...922..253M}. The inferred distance $d\simeq3.3$ kpc, coupled with a Galactic latitude $b=+37\deg$, provide an unusually high height above the Galactic disk ($z\simeq3$ kpc). This supports the idea that  Calvera was born in the Galactic halo, most likely from the explosion of a run-away massive star or, possibly, in a more unusual event involving a halo star, such as, e.g., the accretion induced collapse of a white dwarf.

Zane et al.\ 2011 \citep{2011MNRAS.410.2428Z} reported the presence of diffuse X-ray emission about 13$'$ west of Calvera, with spectral properties consistent with a SNR; recently,  radio \citep{2022A&A...667A..71A} and $\gamma$-ray \citep{2022ApJ...941..194X,2023MNRAS.518.4132A} counterparts of this remnant were discovered. These findings might imply that Calvera is younger than inferred from its timing parameters. The refined measurement of the proper motion (coming soon) could confirm or discard the association between Calvera and this supernova remnant.


The storage of several hundreds of ks of \xmm and \chandra pointings, and the increased sensitivity of the all sky survey in the soft X-ray band provided by \erosita allowed us to discover new INS candidates even in the absence of pulsations.  This could be caused by an unfavorable orientation of the rotation and magnetic axes, or by the intrinsic lack, or faintness, of non-thermal magnetospheric emission. The most promising INS candidates are 
2XMM J104608.7$-$594306 \citep{2009A&A...498..233P,2015A&A...583A.117P},
4XMM J022141.5$-$735632 \citep{2022MNRAS.509.1217R,2022A&A...666A.148P},
eRASSU J065715.3$+$260428 and eRASSU        J131716.9$-$402647 \citep{2023A&A...674A.155K},
but with the data release of the first \erosita catalog we expect a significant enlargement of the sample \citep{2017AN....338..213P,2023IAUS..363..301K}.

\subsection{Long Period Pulsars}\label{sec:longp}

Finally, there is a new emergent class of pulsars characterized by a long period:  PSR J2251$-$3711 \citep[$P = 12.1$ s,][]{2020MNRAS.493.1165M}, PSR J1903$+$0433 \citep[$P = 14.0$ s,][]{2021RAA....21..107H}, PSR J0250$+$5854 \citep[$P = 23.5$ s,][]{2018ApJ...866...54T}, and PSR J0901$-$4046 \citep[$P = 75.9$ s,][]{2022NatAs...6..828C}. They are located near magnetars and \xds on the $P-\pdot$ diagram, therefore they are characterized by high $B\pdx{dip} \sim 10^{13}-10^{14}$ G and long $\tau_c \sim 10^7$ yr, but they are radio sources without any X-ray counterpart \citep{2022ApJ...940...72R,2023MNRAS.520.5960T}. 
An even more exotic object is without any doubts GLEAM-X J162759.5$-$523504.3 (hereafter GLEAM-X), that showed pulsating radio emission with a period of $\sim\!18$ minutes \citep{2022Natur.601..526H}. The period derivative is not measured, yet, preventing robust determination of the source nature, but the upper limit of $\pdot<2\times10^{-9}$ \sps sets an incredibly strong $B\pdx{dip}<3\times10^{16}$.
Differently from the others long period pulsars observed so far, GLEAM-X is a transient source 
with a very variable flux, showing periods of ``radio outburst'' lasting a few months, a 90\% linear polarization, and a very spiky and variable pulse profile.
The properties of GLEAM-X resemble in some sense those of the 6.7-hr pulsating source at the center of SNR RCW 103, see Section~\ref{sec:cco}.

This new (?) class of radio pulsar challenges our comprehension of radio emission mechanism, as they lie well below the ``death line'' obtained in the standard framework of inner vacuum-gap curvature radiation\footnote{Although this tension can be reconciled if a space-charge-limited flow model with a multipolar magnetic field configuration is considered \citep{2022NatAs...6..828C}.}, see Figure~\ref{fig:PPdot},  
and they have a radio luminosity $\gtrsim\!\edot_{\rm rot}$, indicating that the emission is not generated purely by spin down but that an additional source of energy (magnetic?) is needed. The upper limits on the X-ray luminosity confirm that they are rather old neutron stars and do not discard the similarities with the magnetar family \citep{2023MNRAS.520.1872B}.

The location of these pulsars on the $P-\pdot$ diagram is hard to reconcile in the classical magneto-thermal evolution, because such strong magnetic fields would decay in timescales smaller than the inferred $\tau_c$ (for fields $\sim\!10^{15}$ G, $\tau_{\rm Hall}$ can be as small as $\sim\!100$ yr, see Eq.\ \ref{eq:hall}).
Possible solutions are either real ages of these objects are much smaller than their characteristic ages due to large initial spin periods, or some mechanism prevents the crustal magnetic field from decaying in time \citep[see][and refrences therein]{2022ApJ...934..184R}. 
Associations with white dwarfs instead of neutron stars have been considered and discarded \citep{2023MNRAS.520.1872B}, but not because of the spin periods: white dwarfs with $P\lesssim70$ s have been found both in binary systems \citep{2022MNRAS.509L..31P,2023MNRAS.523.3043R} and as isolated sources \citep{2021ApJ...923L...6K}.



\section{Multipolar magnetic field}
In this section I will report the growing evidences that the magnetic field has complicated field lines (multipolar and toroidal components) in the crust and above the surface also in ordinary RPPs: they concern the polar cap size (Section~\ref{sec:pc}), the presence of absorption features in the soft X-ray band (Section~\ref{sec:abs}), and the thermal surface map that can be inferred from timing and spectral analysis  (Section~\ref{sec:thx}).

\subsection{Polar cap size}\label{sec:pc}

When the neutron star has more than one million years, thermal X-rays are produced from hot spots, which are heated by backward-accelerated magnetospheric particles.
The relation between non-thermal and thermal emission mechanisms is of particular interest for the study of particle acceleration in the pulsar magnetosphere \citep{2001ApJ...556..987H,2002ApJ...568..862H,1986ApJ...300..500C,1986ApJ...300..522C}.

In the classical dipolar scenario, the hot spots are identical and antipodal, are situated in correspondence of the magnetic poles and have a size $R\pdx{dip}= \sqrt{2\pi R_*^3/Pc} \approx 145 P\pdx{1\,s}^{-1/2}$~m.
Several \xmm and \chandra observations have been devoted to measure the characteristics of the hot spots of the brightest old pulsars. Currently, there are only eight pulsars older than 1 Myr with a hot spot emission detected with enough confidence, while in the majority of cases the presence of a thermal component is still under debate 
\citep{2018A&A...615A..73R,2018ApJ...865..116I,2018ApJ...869...97A}.  The emitting radii, inferred with blackbody fits, are a few meters only, and not all the pulse profiles are symmetric. 

It was proposed \citep{2003A&A...407..315G,2008ApJ...686..497G} that such a small emitting areas can be explained if we consider that, close to  the star surface, the magnetic field is stronger than the dipolar field and described by multipolar components: 
\begin{equation}
    B\pdx{PC} = B\pdx{dip} \left( \frac{R\pdx{dip}}{R\pdx{PC}} \right)^2 \approx 10^2 B\pdx{dip}
    \label{eq:Bmulti}
\end{equation}
and the conservation of the magnetic flux through the polar cap area explain why we observe $R\pdx{PC} \sim 10-40$ m.
The asymmetry of the pulse profile also denotes that the geometry of the magnetic field close to the surface may be off-centred, and this would imply two caps non-identical, non-antipodal, and out of phase with respect to the radio main peak.
Recent results obtained with \nicer \citep{2019ApJ...887L..26B,2019ApJ...887L..23B,2019ApJ...887L..21R,2023arXiv230903620P} also report evidences for non-dipolar magnetic fields, but they refer to the older MSPs that have $B\pdx{dip}\sim10^9$ G.

These discrepancies can also  be addressed using more realistic thermal models, accounting for the effects of magnetic field on  either a condensed surface or  an atmosphere. It is well known that atmospheric models are harder than blackbodies, thus yielding best-fit temperatures  a factor about two lower and, as a consequence, larger emitting areas. For three pulsars, the latter are consistent with the dipole polar cap (PSR B0950$+$08 \citep{zav04}, PSR B0823$+$26 \citep{her18}, PSR B0943$+$10 \citep{2019ApJ...872...15R}). Moreover, differently from the isotropic blackbody emission, the magnetic field imprints a beaming on the emergent radiation, even if the surface temperature is uniform, and the resulting pulse profiles are naturally non-symmetrical \citep{2019A&A...627A..69R}.
However, these models depend on many more parameters than a simple blackbody, such as the chemical composition, the star mass and radius, the magnetic field, the system geometry, that in most cases are unknown and are difficult to constrain in these rather faint old pulsars.



\subsection{Absorption lines}\label{sec:abs}

Usually, cyclotron features are detected in the (thermal) X-ray spectra of particular classes of INSs (Sections~\ref{sec:mag}, ~\ref{sec:cco}, \ref{sec:xd}), that have the appropriate magnetic field to produce electron or proton absorption lines, respectively.
On the contrary, for the ordinary pulsars (with $B\pdx{dip}\sim10^{12}$~G) neither of this conditions is satisfied at the star surface and hence no analogous cyclotron features are expected in soft X-rays (Eqq.\ \ref{eq:Bcycp} and \ref{eq:Bcyce}).

Nevertheless, in the last few years some ordinary RPPs 
(PSR J1740$+$1000 \citep{2012Sci...337..946K}, PSR B1133$+$16 \citep{2018A&A...615A..73R}, and PSR B0656$+$14 \citep{2018ApJ...869...97A})
showed evidence for the presence of features in their X-ray spectra at about $\sim\!0.5$ keV.  The case of PSR J1740$+$1000 is extremely intriguing because, after a glitch occurred in 2012, the X-ray spectrum has changed and the absorption lines seem to have disappeared \citep{2022MNRAS.513.3113R}.

If the electrons are responsible for such features, they must be located high in the magnetosphere, at several stellar radii above the stellar surface where the dipole field is weaker ($B \propto (r/R)^{-3}$). Thus, the X-ray photons can be produced on the surface or in the magnetosphere. 
Alternatively, if the lines are attributed to protons close to the star surface, then a magnetic field $B = 4.5 \times 10^{13} \mathrm{~G} \approx 10 \times B\pdx{dip}$ is needed. This is not implausible if we consider that, on the surface, the dipole approximation may be no longer valid, but a source of soft X-ray photons on the surface, with an extension appropriate for this context (i.e. $R\pdx{PC}=R\pdx{dip}/\sqrt{10}\approx40$ m, see Eq.\ \ref{eq:Bmulti}), is needed. 

\subsection{Thermal emission}\label{sec:thx}

\begin{figure*}[ht]
\centering
\includegraphics[width=0.49\textwidth]{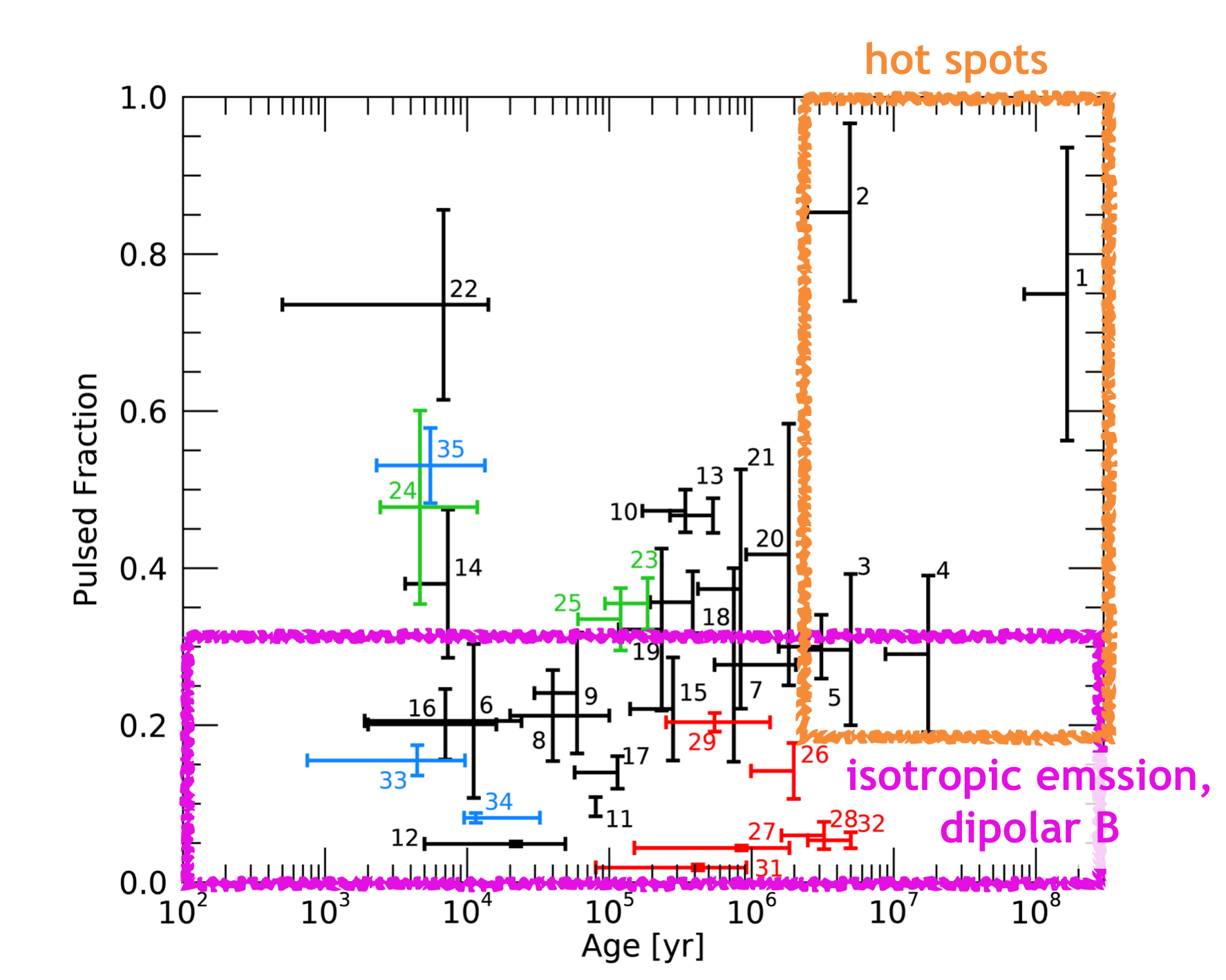}
\includegraphics[width=0.49\textwidth]{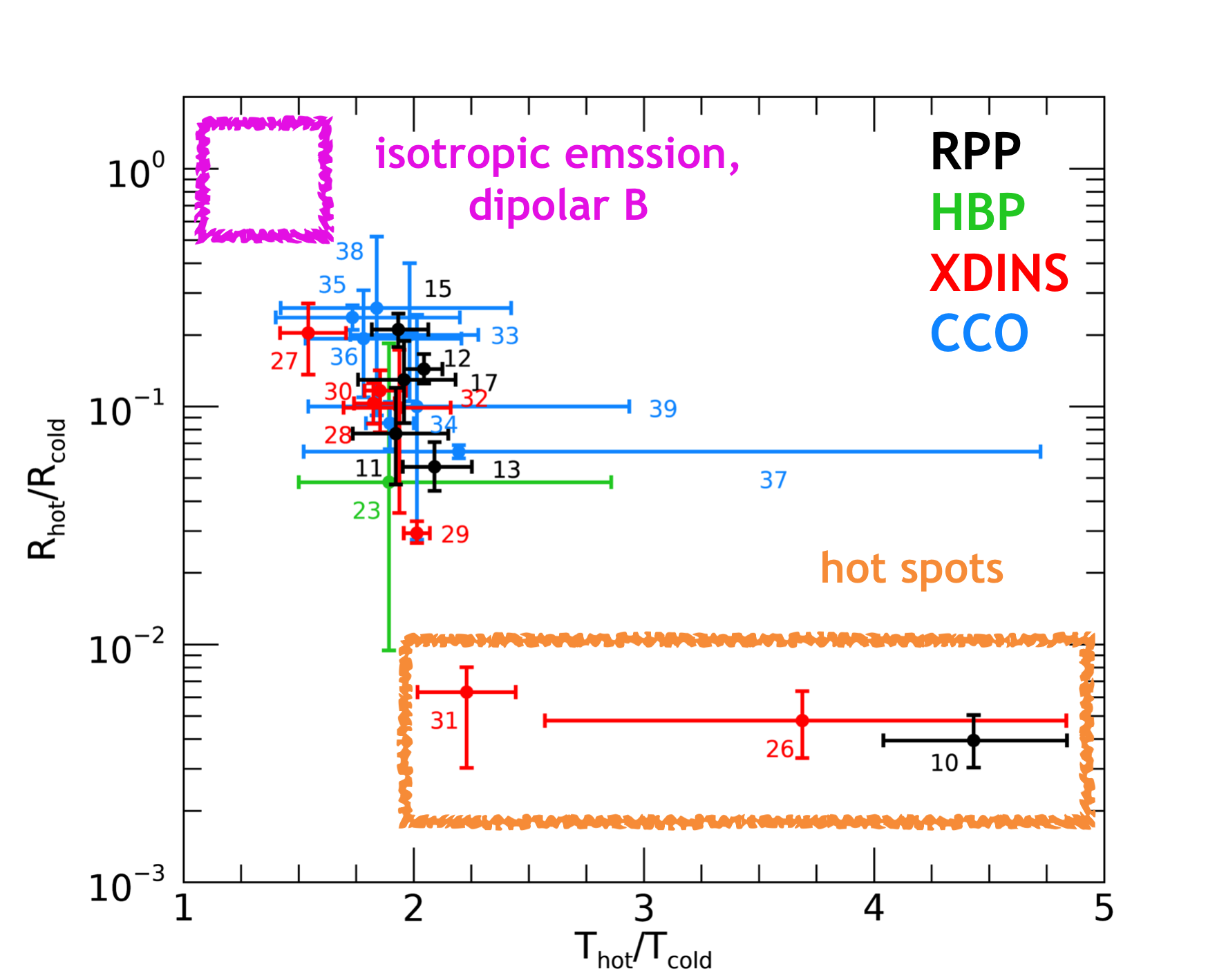}
\caption{\small Pulsed fraction vs age (left panel) and ratio of the two emitting radii vs ratio of the two temperatures derived from  2BB fits of thermal spectra (right panel). The black dots represent RPPs (\#1 PSR J0108$-$1431 \citep{aru19}, \#2 PSR B0823+26 \citep{her18}, \#3 PSR B0943+10 \citep{2019ApJ...872...15R}, \#4 PSR B0950+08 \citep{zav04}, \#5  PSR B1929+10 \citep{mis08},
\#6 PSR J0007+7303 \citep{2010ApJ...725L...6C}, \#7 PSR J0357+3205 \citep{2013ApJ...765...36M}, \#8 PSR J0538+2817 \citep{2007ApJ...654..487N}, \#9 PSR J0633+0632 \citep{2020MNRAS.493.1874D}, \#10 PSR J0633+1746/Geminga \citep{2014ApJ...793...88M}, \#11 PSR B0656+14 \citep{2018ApJ...869...97A}, \#12 PSR B0833$-$45/Vela \citep{2007ApJ...669..570M}, \#13 PSR B1055$-$52 \citep{2022MNRAS.513.3113R}, \#14 PSR J1357$-$6429 \citep{2012ApJ...744...81C}, \#15 1RXS J141256.0$+$792204/Calvera \citep{2021ApJ...922..253M}, \#16 PSR B1706$-$44 \citep{mcg04}, \#17 PSR J1740+1000 \citep{2022MNRAS.513.3113R}, \#18 PSR J1741$-$2054 \citep{2014ApJ...790...51M}, \#19 PSR B1822$-$09 \citep{her17}, \#20 PSR J1836+5925 \citep{2014ApJ...793L...8L}, \#21 PSR J1957+5033 \citep{2021MNRAS.501.4998Z}, \#22 PSR J2021+4026 \citep{2021A&A...646A.117R}),
the green dots represent high-B pulsars (\#23 PSR J0726$-$2612 \citep{2019ApJ...872...15R}, \#24 PSR J1119$-$6127 \citep{2012ApJ...761...65N}, \#25 PSR J1819$-$1458 \citep{mil13}),
the red dots represent XDINSs (\#26 RX J0420.0$-$5022, \#27 RX J0720.4$-$3125, \#28 RX J0806.4$-$4123, \#29 RX J1308.6+2127, \#30 RX J1605.3+3249, \#31 RX J1856.5$-$3754 and \#32 RX J2143.0+0654, \citep{yon19}),
the light blue dots represent CCOs (\#33 RX J0822.0$-$4300 \citep{got10}, \#34 1E 1207.4$-$5209 \citep{big03}, \#35 CXOU J185238.6+004020 \citep{2010ApJ...709..436H}, \#36 1WGA J1713.4$-$3949 \citep{2004A&A...427..199C}; \#37 XMMU J172054.5$-$372652 \citep{2011ApJ...731...70L}; \#38 XMMU J173203.3$-$344518 \citep{2010ApJ...712..790T}, \#39 CXOU J232327.9+584842, \citep{pav09}). Both panels show the overall expected regions for an isotropic emission and a dipolar magnetic field \citep{gre83} (in magenta) and the hot spots from returning currents (in orange).} 
\label{fig:multi}
\end{figure*}

The phase-resolved spectral analysis of the X-ray thermal component can help to interpret the temperature map of the neutron star surface, that is directly influenced by the magnetic field structure. The bulk of thermal energy is stored in the star interior, that is essentially isothermal, and arises from the envelope, that is only 100 m thick and has a density up to $10^{10}$ g cm$^{-3}$. 
In the presence of a strong magnetic field, the electron transport in directions perpendicular to the field is heavily suppressed, while it is enhanced parallel to the field, resulting in an anisotropic heat transport and a non-uniform temperature distribution on the surface \citep[e.g.]{1985MNRAS.213..313H}.

There are two quantities that are directly correlated to the temperature map: the pulsed fraction of the thermal component, and the ratio between the multiple components that are needed to fit the thermal spectrum. Both of these quantities are represented in Figure~\ref{fig:multi} for all the thermally emitting RPPs (black), high-B pulsars (green), \xds (red) and CCOs (light blue) known.

The left panel of Figure~\ref{fig:multi} shows the pulsed fraction as a function of the INS age. There is no correlation between the PF and the age, nor between the INS class. While for the older ($>\!10^6$ yr) RPPs the high pulsed fraction can be explained with the small size of the emitting area and a favorable orientation of $\Omega$, $\mu$ and the LOS (orange area), there are at least four objects younger than $10^4$ yr (PSR J2021$+$4026 \citep{2021A&A...646A.117R}, PSR J1119$-$6127 \citep{2012ApJ...761...65N}, PSR J1357$+$6429 \citep{2012ApJ...744...81C} and PSR J1852+0040 \citep{2010ApJ...709..436H}) for which we expect that the thermal emission comes from a large fraction of the cooling surface. 
In case of isotropic emission, several authors \citep[e.g.]{2019A&A...627A..69R} showed that the maximum PF expected should be $\lesssim\!0.3$ (magenta area), even if the emitting area is smaller than the entire surface \citep{tur13} and even if the temperature distribution is influenced by a dipolar magnetic field \citep{gre83}.
We need a non-isotropic emission mechanism, as in the case of magnetized atmosphere, or a highly inhomogeneous temperature distribution, that can be produced only with multipolar component and/or toroidal component \citep{per06a,per06b,gep06}.

The right panel of Figure~\ref{fig:multi} shows instead all the INSs whose thermal spectrum can be fitted with, at least, two blackbodies, that mimic the two extremes of a more complex temperature distribution. Therefore in this plot all the old RPPs are excluded because the surface has cooled down and it is not longer visible in the X-rays. The figure shows the ratios of emitting radii ($R\pdx{hot}/R\pdx{cold}$) and temperatures ($T\pdx{hot}/T\pdx{cold}$), values that are model and distance independent. 

It is very striking to observe that, for all these INSs except three (Geminga \citep{2014ApJ...793...88M}, RX J0420.0$-$5022 and \rxj [Section~\ref{sec:rxj}]), the ratio $R\pdx{hot}/R\pdx{cold}$ is in the range $0.03-0.3$. Thus it is difficult to interpret the hotter and colder components in terms of emission from hot spots and from  the whole surface ($R\lesssim200$ m and $R\sim12-16$ km, respectively, see orange area).  
We also note that all different INS classes are placed in the same area of this space parameter: this means that, despite the differences, they have a similar thermal map. This is indeed quite surprising since the surface temperature is expected to evolve in time and to be sensitive to the initial magnetic field configuration in the star crust (see \citep{2019LRCA....5....3P} for a review). 
Yakovlev 2021 \citep{2021MNRAS.506.4593Y} and Rigoselli et al.\ 2022 \citep{2022MNRAS.513.3113R} showed that the simplest case of a blackbody emission with a temperature distribution resulting from a dipolar magnetic field \citep{gre83} gives values of the $R\pdx{hot}/R\pdx{cold}$ and $T\pdx{hot}/T\pdx{cold}$ ratios clearly inconsistent with the observed ones (magenta area).

\section{Final remarks}

Recent findings show that the \xd class is probably more variegate than previously thought: two of the softer objects  show non-thermal emission, that could have a magnetospheric origin as for the ordinary pulsars, while the absence of radio emission that was the main difference between \xds and RPPs could be explained by orientation effects.

The analysis of timing and spectral X-ray data of different classes of INSs support the mounting evidence that ordinary RPPs may have complicated thermal maps and magnetic fields capable to induce cyclotron absorption in the keV energy range, as already seen for \xds and magnetars in quiescence. The latter were successfully interpreted with 3D magneto-thermal simulations of strong and tilted toroidal magnetic fields \citep{2020ApJ...903...40D,2021ApJ...914..118D,2021NatAs...5..145I}.

It would be interesting to apply similar models also to other classes of INSs and, even more, to investigate the possible cause of the observed clustering in temperatures and radii ratios, that cover sources with a large spread of ages and magnetic field strengths.


\footnotesize
\bibliographystyle{JHEP}
\bibliography{bibliography.bib}

\providecommand{\href}[2]{#2}\begingroup\raggedright\begin{thebibliography}{100}

\bibitem{hew68}
{Hewish}, A., {Bell}, S.~J., {Pilkington}, J.~D.~H., {Scott}, P.~F., \&
  {Collins}, R.~A., \,\href{https://doi.org/10.1038/217709a0}{\emph{\nat}
  {\bfseries 217} (1968) 709}.

\bibitem{2005AJ....129.1993M}
{Manchester}, R.~N., {Hobbs}, G.~B., {Teoh}, A., \& {Hobbs}, M.,
  \,\href{https://doi.org/10.1086/428488}{\emph{\aj} {\bfseries 129} (2005)
  1993}.

\bibitem{2013FrPhy...8..679H}
{Harding}, A.~K.,
  \,\href{https://doi.org/10.1007/s11467-013-0285-0}{\emph{Frontiers of
  Physics} {\bfseries 8} (2013) 679}.

\bibitem{2014AN....335..262I}
{Igoshev}, A.~P., {Popov}, S.~B., \& {Turolla}, R.,
  \,\href{https://doi.org/10.1002/asna.201312029}{\emph{Astronomische
  Nachrichten} {\bfseries 335} (2014) 262}.

\bibitem{2018IAUS..337....3K}
{Kaspi}, V.~M., \, in \emph{Pulsar Astrophysics the Next Fifty Years},
  {Weltevrede}, P., {Perera}, B.~B.~P., {Preston}, L.~L., \& {Sanidas}, S.,
  eds., vol.~337, (2018), pp.~3--8,
  \href{https://doi.org/10.1017/S1743921317010390}{DOI:10.1017/S1743921317010390}.

\bibitem{2023Univ....9..273P}
{Popov}, S.~B.,
  \,\href{https://doi.org/10.3390/universe9060273}{\emph{Universe} {\bfseries
  9} (2023) 273}.

\bibitem{bha92}
{Bhattacharya}, D., {Wijers}, R.~A.~M.~J., {Hartman}, J.~W., \& {Verbunt}, F.,
  \,{\emph{\aap} {\bfseries 254} (1992) 198}.

\bibitem{2023NatAs.tmp..208H}
{H.~E.~S.~S. Collaboration}, {Aharonian}, F., {Ait Benkhali}, F., {et~al.},
  \,\href{https://doi.org/10.1038/s41550-023-02052-3}{\emph{Nature Astronomy}
  (2023) }.

\bibitem{pot20}
{Potekhin}, A.~Y., {Zyuzin}, D.~A., {Yakovlev}, D.~G., {Beznogov}, M.~V., \&
  {Shibanov}, Y.~A.,
  \,\href{https://doi.org/10.1093/mnras/staa1871}{\emph{\mnras} {\bfseries 496}
  (2020) 5052}.

\bibitem{bar98}
{Baring}, M.~G. \& {Harding}, A.~K.,
  \,\href{https://doi.org/10.1086/311679}{\emph{\apjl} {\bfseries 507} (1998)
  L55}.

\bibitem{bar01}
{Baring}, M.~G. \& {Harding}, A.~K.,
  \,\href{https://doi.org/10.1086/318390}{\emph{\apj} {\bfseries 547} (2001)
  929}.

\bibitem{2023RNAAS...7..213H}
{Halder}, P., {Goswami}, S., {Halder}, P., {Ghosh}, U., \& {Konar}, S.,
  \,\href{https://doi.org/10.3847/2515-5172/ad00ac}{\emph{RNAAS} {\bfseries 7}
  (2023) 213}.

\bibitem{man17}
{Manchester}, R.~N.,
  \,\href{https://doi.org/10.1007/s12036-017-9469-2}{\emph{JApA} {\bfseries 38}
  (2017) 42}.

\bibitem{ng11}
{Ng}, C.~Y. \& {Kaspi}, V.~M., \, in \emph{American Institute of Physics
  Conference Series}, {G{\"o}{\u{g}}{\"u}{\textcommabelow s}}, E., {Belloni},
  T., \& {Ertan}, {\"U}., eds., vol.~1379, (2011), pp.~60--69,
  \href{https://doi.org/10.1063/1.3629486}{DOI:10.1063/1.3629486}.

\bibitem{2014ApJS..212....6O}
{Olausen}, S.~A. \& {Kaspi}, V.~M.,
  \,\href{https://doi.org/10.1088/0067-0049/212/1/6}{\emph{\apjs} {\bfseries
  212} (2014) 6}.

\bibitem{2018MNRAS.474..961C}
{Coti Zelati}, F., {Rea}, N., {Pons}, J.~A., {Campana}, S., \& {Esposito}, P.,
  \,\href{https://doi.org/10.1093/mnras/stx2679}{\emph{\mnras} {\bfseries 474}
  (2018) 961}.

\bibitem{DT92}
{Duncan}, R.~C. \& {Thompson}, C.,
  \,\href{https://doi.org/10.1086/186413}{\emph{\apjl} {\bfseries 392} (1992)
  L9}.

\bibitem{DT95}
{Thompson}, C. \& {Duncan}, R.~C.,
  \,\href{https://doi.org/10.1093/mnras/275.2.255}{\emph{\mnras} {\bfseries
  275} (1995) 255}.

\bibitem{DT96}
{Thompson}, C. \& {Duncan}, R.~C.,
  \,\href{https://doi.org/10.1086/178147}{\emph{\apj} {\bfseries 473} (1996)
  322}.

\bibitem{2017ARA&A..55..261K}
{Kaspi}, V.~M. \& {Beloborodov}, A.~M.,
  \,\href{https://doi.org/10.1146/annurev-astro-081915-023329}{\emph{\araa}
  {\bfseries 55} (2017) 261}.

\bibitem{2021ASSL..461...97E}
{Esposito}, P., {Rea}, N., \& {Israel}, G.~L., \, in \emph{Timing Neutron
  Stars: Pulsations, Oscillations and Explosions}, {Belloni}, T.~M.,
  {M{\'e}ndez}, M., \& {Zhang}, C., eds., vol.~461 of \emph{Astrophysics and
  Space Science Library}, (2021), pp.~97--142,
  \href{https://doi.org/10.1007/978-3-662-62110-3\_3}{DOI:10.1007/978-3-662-62110-3\_3}.

\bibitem{2002ApJ...574..332T}
{Thompson}, C., {Lyutikov}, M., \& {Kulkarni}, S.~R.,
  \,\href{https://doi.org/10.1086/340586}{\emph{\apj} {\bfseries 574} (2002)
  332}.

\bibitem{2003ApJ...586L..65R}
{Rea}, N., {Israel}, G.~L., {Stella}, L., {et~al.},
  \,\href{https://doi.org/10.1086/374585}{\emph{\apjl} {\bfseries 586} (2003)
  L65}.

\bibitem{2014AN....335..274T}
{Tiengo}, A., {Esposito}, P., {Mereghetti}, S., {et~al.},
  \,\href{https://doi.org/10.1002/asna.201312031}{\emph{Astronomische
  Nachrichten} {\bfseries 335} (2014) 274}.

\bibitem{2022Sci...378..646T}
{Taverna}, R., {Turolla}, R., {Muleri}, F., {et~al.},
  \,\href{https://doi.org/10.1126/science.add0080}{\emph{Science} {\bfseries
  378} (2022) 646}.

\bibitem{2023ApJ...944L..27Z}
{Zane}, S., {Taverna}, R., {Gonz{\'a}lez-Caniulef}, D., {et~al.},
  \,\href{https://doi.org/10.3847/2041-8213/acb703}{\emph{\apjl} {\bfseries
  944} (2023) L27}.

\bibitem{2023ApJ...954...88T}
{Turolla}, R., {Taverna}, R., {Israel}, G.~L., {et~al.},
  \,\href{https://doi.org/10.3847/1538-4357/aced05}{\emph{\apj} {\bfseries 954}
  (2023) 88}.

\bibitem{cam07}
{Camilo}, F., {Reynolds}, J., {Johnston}, S., {et~al.},
  \,\href{https://doi.org/10.1086/516630}{\emph{\apjl} {\bfseries 659} (2007)
  L37}.

\bibitem{cam08}
{Camilo}, F., {Reynolds}, J., {Johnston}, S., {Halpern}, J.~P., \& {Ransom},
  S.~M., \,\href{https://doi.org/10.1086/587054}{\emph{\apj} {\bfseries 679}
  (2008) 681}.

\bibitem{2008Sci...319.1802G}
{Gavriil}, F.~P., {Gonzalez}, M.~E., {Gotthelf}, E.~V., {et~al.},
  \,\href{https://doi.org/10.1126/science.1153465}{\emph{Science} {\bfseries
  319} (2008) 1802}.

\bibitem{2016ApJ...829L..21A}
{Archibald}, R.~F., {Kaspi}, V.~M., {Tendulkar}, S.~P., \& {Scholz}, P.,
  \,\href{https://doi.org/10.3847/2041-8205/829/1/L21}{\emph{\apjl} {\bfseries
  829} (2016) L21}.

\bibitem{2010MNRAS.405.1787E}
{Esposito}, P., {Israel}, G.~L., {Turolla}, R., {et~al.},
  \,\href{https://doi.org/10.1111/j.1365-2966.2010.16551.x}{\emph{\mnras}
  {\bfseries 405} (2010) 1787}.

\bibitem{2012ApJ...754...27R}
{Rea}, N., {Israel}, G.~L., {Esposito}, P., {et~al.},
  \,\href{https://doi.org/10.1088/0004-637X/754/1/27}{\emph{\apj} {\bfseries
  754} (2012) 27}.

\bibitem{2014ApJ...781L..17R}
{Rea}, N., {Vigan{\`o}}, D., {Israel}, G.~L., {Pons}, J.~A., \& {Torres},
  D.~F., \,\href{https://doi.org/10.1088/2041-8205/781/1/L17}{\emph{\apjl}
  {\bfseries 781} (2014) L17}.

\bibitem{2017hsn..book.1401B}
{Bisnovatyi-Kogan}, G.~S., \emph{{Young Neutron Stars with Soft Gamma Ray
  Emission and Anomalous X-Ray Pulsars}},  in \emph{Handbook of Supernovae},
  {Alsabti}, A.~W. \& {Murdin}, P., eds., p.~1401 (2017),
  \href{https://doi.org/10.1007/978-3-319-21846-5\_70}{DOI:10.1007/978-3-319-21846-5\_70}.

\bibitem{2015MNRAS.446.1121G}
{Gourgouliatos}, K.~N. \& {Cumming}, A.,
  \,\href{https://doi.org/10.1093/mnras/stu2140}{\emph{\mnras} {\bfseries 446}
  (2015) 1121}.

\bibitem{2015MNRAS.449.2047L}
{Lander}, S.~K., {Andersson}, N., {Antonopoulou}, D., \& {Watts}, A.~L.,
  \,\href{https://doi.org/10.1093/mnras/stv432}{\emph{\mnras} {\bfseries 449}
  (2015) 2047}.

\bibitem{2011ApJ...740..105T}
{Turolla}, R., {Zane}, S., {Pons}, J.~A., {Esposito}, P., \& {Rea}, N.,
  \,\href{https://doi.org/10.1088/0004-637X/740/2/105}{\emph{\apj} {\bfseries
  740} (2011) 105}.

\bibitem{mig19}
{Mignani}, R.~P., {De Luca}, A., {Zharikov}, S., {et~al.},
  \,\href{https://doi.org/10.1093/mnras/stz1195}{\emph{\mnras} {\bfseries 486}
  (2019) 5716}.

\bibitem{2017JPhCS.932a2006D}
{De Luca}, A., \, in \emph{Journal of Physics Conference Series}, vol.~932 of
  \emph{Journal of Physics Conference Series}, (2017), p.~012006,
  \href{https://doi.org/10.1088/1742-6596/932/1/012006}{DOI:10.1088/1742-6596/932/1/012006}.

\bibitem{zav00}
{Zavlin}, V.~E., {Pavlov}, G.~G., {Sanwal}, D., \& {Tr{\"u}mper}, J.,
  \,\href{https://doi.org/10.1086/312866}{\emph{\apjl} {\bfseries 540} (2000)
  L25}.

\bibitem{got05}
{Gotthelf}, E.~V., {Halpern}, J.~P., \& {Seward}, F.~D.,
  \,\href{https://doi.org/10.1086/430300}{\emph{\apj} {\bfseries 627} (2005)
  390}.

\bibitem{got09}
{Gotthelf}, E.~V. \& {Halpern}, J.~P.,
  \,\href{https://doi.org/10.1088/0004-637X/695/1/L35}{\emph{\apjl} {\bfseries
  695} (2009) L35}.

\bibitem{big03}
{Bignami}, G.~F., {Caraveo}, P.~A., {De Luca}, A., \& {Mereghetti}, S.,
  \,\href{https://doi.org/10.1038/nature01703}{\emph{\nat} {\bfseries 423}
  (2003) 725}.

\bibitem{got10}
{Gotthelf}, E.~V., {Perna}, R., \& {Halpern}, J.~P.,
  \,\href{https://doi.org/10.1088/0004-637X/724/2/1316}{\emph{\apj} {\bfseries
  724} (2010) 1316}.

\bibitem{hal10}
{Halpern}, J.~P. \& {Gotthelf}, E.~V.,
  \,\href{https://doi.org/10.1088/0004-637X/709/1/436}{\emph{\apj} {\bfseries
  709} (2010) 436}.

\bibitem{got13}
{Gotthelf}, E.~V., {Halpern}, J.~P., \& {Alford}, J.,
  \,\href{https://doi.org/10.1088/0004-637X/765/1/58}{\emph{\apj} {\bfseries
  765} (2013) 58}.

\bibitem{sha12}
{Shabaltas}, N. \& {Lai}, D.,
  \,\href{https://doi.org/10.1088/0004-637X/748/2/148}{\emph{\apj} {\bfseries
  748} (2012) 148}.

\bibitem{2020MNRAS.495.1692G}
{Gourgouliatos}, K.~N., {Hollerbach}, R., \& {Igoshev}, A.~P.,
  \,\href{https://doi.org/10.1093/mnras/staa1295}{\emph{\mnras} {\bfseries 495}
  (2020) 1692}.

\bibitem{2016ApJ...828L..13R}
{Rea}, N., {Borghese}, A., {Esposito}, P., {et~al.},
  \,\href{https://doi.org/10.3847/2041-8205/828/1/L13}{\emph{\apjl} {\bfseries
  828} (2016) L13}.

\bibitem{2016MNRAS.463.2394D}
{D'A{\`\i}}, A., {Evans}, P.~A., {Burrows}, D.~N., {et~al.},
  \,\href{https://doi.org/10.1093/mnras/stw2023}{\emph{\mnras} {\bfseries 463}
  (2016) 2394}.

\bibitem{2011ApJ...736..117K}
{Kaplan}, D.~L., {Kamble}, A., {van Kerkwijk}, M.~H., \& {Ho}, W.~C.~G.,
  \,\href{https://doi.org/10.1088/0004-637X/736/2/117}{\emph{\apj} {\bfseries
  736} (2011) 117}.

\bibitem{2009ApJ...702..692K}
{Kondratiev}, V.~I., {McLaughlin}, M.~A., {Lorimer}, D.~R., {et~al.},
  \,\href{https://doi.org/10.1088/0004-637X/702/1/692}{\emph{\apj} {\bfseries
  702} (2009) 692}.

\bibitem{2007Ap&SS.308..191V}
{van Kerkwijk}, M.~H. \& {Kaplan}, D.~L.,
  \,\href{https://doi.org/10.1007/s10509-007-9343-9}{\emph{\apss} {\bfseries
  308} (2007) 191}.

\bibitem{2009ASSL..357..141T}
{Turolla}, R., \, in \emph{Astrophysics and Space Science Library}, {Becker},
  W., ed., vol.~357 of \emph{Astrophysics and Space Science Library}, (2009),
  p.~141,
  \href{https://doi.org/10.1007/978-3-540-76965-1\_7}{DOI:10.1007/978-3-540-76965-1\_7}.

\bibitem{2003A&A...403L..19H}
{Haberl}, F., {Schwope}, A.~D., {Hambaryan}, V., {Hasinger}, G., \& {Motch},
  C., \,\href{https://doi.org/10.1051/0004-6361:20030450}{\emph{\aap}
  {\bfseries 403} (2003) L19}.

\bibitem{2004ApJ...608..432V}
{van Kerkwijk}, M.~H., {Kaplan}, D.~L., {Durant}, M., {Kulkarni}, S.~R., \&
  {Paerels}, F., \,\href{https://doi.org/10.1086/386299}{\emph{\apj} {\bfseries
  608} (2004) 432}.

\bibitem{2004A&A...424..635H}
{Haberl}, F., {Motch}, C., {Zavlin}, V.~E., {et~al.},
  \,\href{https://doi.org/10.1051/0004-6361:20040440}{\emph{\aap} {\bfseries
  424} (2004) 635}.

\bibitem{2005ApJ...627..397Z}
{Zane}, S., {Cropper}, M., {Turolla}, R., {et~al.},
  \,\href{https://doi.org/10.1086/430138}{\emph{\apj} {\bfseries 627} (2005)
  397}.

\bibitem{2015ApJ...807L..20B}
{Borghese}, A., {Rea}, N., {Coti Zelati}, F., {Tiengo}, A., \& {Turolla}, R.,
  \,\href{https://doi.org/10.1088/2041-8205/807/1/L20}{\emph{\apjl} {\bfseries
  807} (2015) L20}.

\bibitem{2017MNRAS.468.2975B}
{Borghese}, A., {Rea}, N., {Coti Zelati}, F., {et~al.},
  \,\href{https://doi.org/10.1093/mnras/stx632}{\emph{\mnras} {\bfseries 468}
  (2017) 2975}.

\bibitem{2008AIPC..968..129K}
{Kaplan}, D.~L., \, in \emph{Astrophysics of Compact Objects}, {Yuan}, Y.-F.,
  {Li}, X.-D., \& {Lai}, D., eds., vol.~968, (2008), pp.~129--136,
  \href{https://doi.org/10.1063/1.2840384}{DOI:10.1063/1.2840384}.

\bibitem{2020MNRAS.497.2883K}
{Kondratyev}, I.~A., {Moiseenko}, S.~G., {Bisnovatyi-Kogan}, G.~S., \&
  {Glushikhina}, M.~V.,
  \,\href{https://doi.org/10.1093/mnras/staa2154}{\emph{\mnras} {\bfseries 497}
  (2020) 2883}.

\bibitem{2021A&A...646A.117R}
{Rigoselli}, M., {Mereghetti}, S., {Taverna}, R., {Turolla}, R., \& {De
  Grandis}, D.,
  \,\href{https://doi.org/10.1051/0004-6361/202039774}{\emph{\aap} {\bfseries
  646} (2021) A117}.

\bibitem{hol02}
{Hollerbach}, R. \& {R{\"u}diger}, G.,
  \,\href{https://doi.org/10.1046/j.1365-8711.2002.05905.x}{\emph{\mnras}
  {\bfseries 337} (2002) 216}.

\bibitem{hol04}
{Hollerbach}, R. \& {R{\"u}diger}, G.,
  \,\href{https://doi.org/10.1111/j.1365-2966.2004.07307.x}{\emph{\mnras}
  {\bfseries 347} (2004) 1273}.

\bibitem{cum04}
{Cumming}, A., {Arras}, P., \& {Zweibel}, E.,
  \,\href{https://doi.org/10.1086/421324}{\emph{\apj} {\bfseries 609} (2004)
  999}.

\bibitem{gon10}
{Gonzalez}, D. \& {Reisenegger}, A.,
  \,\href{https://doi.org/10.1051/0004-6361/201015084}{\emph{\aap} {\bfseries
  522} (2010) A16}.

\bibitem{2013MNRAS.434..123V}
{Vigan{\`o}}, D., {Rea}, N., {Pons}, J.~A., {et~al.},
  \,\href{https://doi.org/10.1093/mnras/stt1008}{\emph{\mnras} {\bfseries 434}
  (2013) 123}.

\bibitem{2014MNRAS.444.1066I}
{Igoshev}, A.~P. \& {Popov}, S.~B.,
  \,\href{https://doi.org/10.1093/mnras/stu1496}{\emph{\mnras} {\bfseries 444}
  (2014) 1066}.

\bibitem{1992A&A...254..198B}
{Bhattacharya}, D., {Wijers}, R. A.~M.~J., {Hartman}, J.~W., \& {Verbunt}, F.,
  \,{\emph{\aap} {\bfseries 254} (1992) 198}.

\bibitem{2001A&A...376..543T}
{Tauris}, T.~M. \& {Konar}, S.,
  \,\href{https://doi.org/10.1051/0004-6361:20010988}{\emph{\aap} {\bfseries
  376} (2001) 543}.

\bibitem{2007A&A...470..303P}
{Pons}, J.~A. \& {Geppert}, U.,
  \,\href{https://doi.org/10.1051/0004-6361:20077456}{\emph{\aap} {\bfseries
  470} (2007) 303}.

\bibitem{2011ApJ...743..183S}
{Speagle}, J.~S., {Kaplan}, D.~L., \& {van Kerkwijk}, M.~H.,
  \,\href{https://doi.org/10.1088/0004-637X/743/2/183}{\emph{\apj} {\bfseries
  743} (2011) 183}.

\bibitem{2019A&A...627A..69R}
{Rigoselli}, M., {Mereghetti}, S., {Suleimanov}, V., {et~al.},
  \,\href{https://doi.org/10.1051/0004-6361/201935485}{\emph{\aap} {\bfseries
  627} (2019) A69}.

\bibitem{2011A&A...534A..74H}
{Hambaryan}, V., {Suleimanov}, V., {Schwope}, A.~D., {et~al.},
  \,\href{https://doi.org/10.1051/0004-6361/201117548}{\emph{\aap} {\bfseries
  534} (2011) A74}.

\bibitem{2017A&A...601A.108H}
{Hambaryan}, V., {Suleimanov}, V., {Haberl}, F., {et~al.},
  \,\href{https://doi.org/10.1051/0004-6361/201630368}{\emph{\aap} {\bfseries
  601} (2017) A108}.

\bibitem{2022MNRAS.516.4932D}
{De Grandis}, D., {Rigoselli}, M., {Mereghetti}, S., {et~al.},
  \,\href{https://doi.org/10.1093/mnras/stac2587}{\emph{\mnras} {\bfseries 516}
  (2022) 4932}.

\bibitem{1996Natur.379..233W}
{Walter}, F.~M., {Wolk}, S.~J., \& {Neuh{\"a}user}, R.,
  \,\href{https://doi.org/10.1038/379233a0}{\emph{\nat} {\bfseries 379} (1996)
  233}.

\bibitem{2010ApJ...724..669W}
{Walter}, F.~M., {Eisenbei{\ss}}, T., {Lattimer}, J.~M., {et~al.},
  \,\href{https://doi.org/10.1088/0004-637X/724/1/669}{\emph{\apj} {\bfseries
  724} (2010) 669}.

\bibitem{2007ApJ...657L.101T}
{Tiengo}, A. \& {Mereghetti}, S.,
  \,\href{https://doi.org/10.1086/513143}{\emph{\apjl} {\bfseries 657} (2007)
  L101}.

\bibitem{2008ApJ...673L.163V}
{van Kerkwijk}, M.~H. \& {Kaplan, D.~L.},
  \,\href{https://doi.org/10.1086/528796}{\emph{\apjl} {\bfseries 673} (2008)
  L163}.

\bibitem{1997A&A...318L..43N}
{Neuhaeuser}, R., {Thomas}, H.~C., {Danner}, R., {Peschke}, S., \& {Walter},
  F.~M.,
  \,\href{https://doi.org/https://ui.adsabs.harvard.edu/abs/1997A&A...318L..43N}{\emph{\aap}
  {\bfseries 318} (1997) L43}.

\bibitem{2017PASJ...69...50Y}
{Yoneyama}, T., {Hayashida}, K., {Nakajima}, H., {Inoue}, S., \& {Tsunemi}, H.,
  \,\href{https://doi.org/10.1093/pasj/psx025}{\emph{\pasj} {\bfseries 69}
  (2017) 50}.

\bibitem{2020ApJ...904...42D}
{Dessert}, C., {Foster}, J.~W., \& {Safdi}, B.~R.,
  \,\href{https://doi.org/10.3847/1538-4357/abb4ea}{\emph{\apj} {\bfseries 904}
  (2020) 42}.

\bibitem{2002A&A...387..993P}
{Possenti}, A., {Cerutti}, R., {Colpi}, M., \& {Mereghetti}, S.,
  \,\href{https://doi.org/10.1051/0004-6361:20020472}{\emph{\aap} {\bfseries
  387} (2002) 993}.

\bibitem{2008ApJ...672.1137R}
{Rutledge}, R.~E., {Fox}, D.~B., \& {Shevchuk}, A.~H.,
  \,\href{https://doi.org/10.1086/522667}{\emph{\apj} {\bfseries 672} (2008)
  1137}.

\bibitem{2011MNRAS.410.2428Z}
{Zane}, S., {Haberl}, F., {Israel}, G.~L., {et~al.},
  \,\href{https://doi.org/10.1111/j.1365-2966.2010.17619.x}{\emph{\mnras}
  {\bfseries 410} (2011) 2428}.

\bibitem{2013ApJ...778..120H}
{Halpern}, J.~P., {Bogdanov}, S., \& {Gotthelf}, E.~V.,
  \,\href{https://doi.org/10.1088/0004-637X/778/2/120}{\emph{\apj} {\bfseries
  778} (2013) 120}.

\bibitem{2021ApJ...922..253M}
{Mereghetti}, S., {Rigoselli}, M., {Taverna}, R., {et~al.},
  \,\href{https://doi.org/10.3847/1538-4357/ac34f2}{\emph{\apj} {\bfseries 922}
  (2021) 253}.

\bibitem{2022A&A...667A..71A}
{Arias}, M., {Botteon}, A., {Bassa}, C.~G., {et~al.},
  \,\href{https://doi.org/10.1051/0004-6361/202244369}{\emph{\aap} {\bfseries
  667} (2022) A71}.

\bibitem{2022ApJ...941..194X}
{Xin}, Y. \& {Guo}, X.,
  \,\href{https://doi.org/10.3847/1538-4357/aca473}{\emph{\apj} {\bfseries 941}
  (2022) 194}.

\bibitem{2023MNRAS.518.4132A}
{Araya}, M., \,\href{https://doi.org/10.1093/mnras/stac3337}{\emph{\mnras}
  {\bfseries 518} (2023) 4132}.

\bibitem{2009A&A...498..233P}
{Pires}, A.~M., {Motch}, C., {Turolla}, R., {Treves}, A., \& {Popov}, S.~B.,
  \,\href{https://doi.org/10.1051/0004-6361/200810966}{\emph{\aap} {\bfseries
  498} (2009) 233}.

\bibitem{2015A&A...583A.117P}
{Pires}, A.~M., {Motch}, C., {Turolla}, R., {et~al.},
  \,\href{https://doi.org/10.1051/0004-6361/201526436}{\emph{\aap} {\bfseries
  583} (2015) A117}.

\bibitem{2022MNRAS.509.1217R}
{Rigoselli}, M., {Mereghetti}, S., \& {Tresoldi}, C.,
  \,\href{https://doi.org/10.1093/mnras/stab2974}{\emph{\mnras} {\bfseries 509}
  (2022) 1217}.

\bibitem{2022A&A...666A.148P}
{Pires}, A.~M., {Motch}, C., {Kurpas}, J., {et~al.},
  \,\href{https://doi.org/10.1051/0004-6361/202244514}{\emph{\aap} {\bfseries
  666} (2022) A148}.

\bibitem{2023A&A...674A.155K}
{Kurpas}, J., {Schwope}, A.~D., {Pires}, A.~M., {Haberl}, F., \& {Buckley},
  D.~A.~H., \,\href{https://doi.org/10.1051/0004-6361/202346375}{\emph{\aap}
  {\bfseries 674} (2023) A155}.

\bibitem{2017AN....338..213P}
{Pires}, A.~M., {Schwope}, A.~D., \& {Motch}, C.,
  \,\href{https://doi.org/10.1002/asna.201713333}{\emph{Astronomische
  Nachrichten} {\bfseries 338} (2017) 213}.

\bibitem{2023IAUS..363..301K}
{Khokhriakova}, A.~D., {Biryukov}, A.~V., \& {Popov}, S.~B., \, in
  \emph{Neutron Star Astrophysics at the Crossroads: Magnetars and the
  Multimessenger Revolution}, {Troja}, E. \& {Baring}, M.~G., eds., vol.~363,
  (2023), pp.~301--304,
  \href{https://doi.org/10.1017/S1743921322000291}{DOI:10.1017/S1743921322000291}.

\bibitem{2020MNRAS.493.1165M}
{Morello}, V., {Keane}, E.~F., {Enoto}, T., {et~al.},
  \,\href{https://doi.org/10.1093/mnras/staa321}{\emph{\mnras} {\bfseries 493}
  (2020) 1165}.

\bibitem{2021RAA....21..107H}
{Han}, J.~L., {Wang}, C., {Wang}, P.~F., {et~al.},
  \,\href{https://doi.org/10.1088/1674-4527/21/5/107}{\emph{RAA} {\bfseries 21}
  (2021) 107}.

\bibitem{2018ApJ...866...54T}
{Tan}, C.~M., {Bassa}, C.~G., {Cooper}, S., {et~al.},
  \,\href{https://doi.org/10.3847/1538-4357/aade88}{\emph{\apj} {\bfseries 866}
  (2018) 54}.

\bibitem{2022NatAs...6..828C}
{Caleb}, M., {Heywood}, I., {Rajwade}, K., {et~al.},
  \,\href{https://doi.org/10.1038/s41550-022-01688-x}{\emph{Nature Astronomy}
  {\bfseries 6} (2022) 828}.

\bibitem{2022ApJ...940...72R}
{Rea}, N., {Coti Zelati}, F., {Dehman}, C., {et~al.},
  \,\href{https://doi.org/10.3847/1538-4357/ac97ea}{\emph{\apj} {\bfseries 940}
  (2022) 72}.

\bibitem{2023MNRAS.520.5960T}
{Tan}, C.~M., {Rigoselli}, M., {Esposito}, P., \& {Stappers}, B.~W.,
  \,\href{https://doi.org/10.1093/mnras/stad492}{\emph{\mnras} {\bfseries 520}
  (2023) 5960}.

\bibitem{2022Natur.601..526H}
{Hurley-Walker}, N., {Zhang}, X., {Bahramian}, A., {et~al.},
  \,\href{https://doi.org/10.1038/s41586-021-04272-x}{\emph{\nat} {\bfseries
  601} (2022) 526}.

\bibitem{2023MNRAS.520.1872B}
{Beniamini}, P., {Wadiasingh}, Z., {Hare}, J., {et~al.},
  \,\href{https://doi.org/10.1093/mnras/stad208}{\emph{\mnras} {\bfseries 520}
  (2023) 1872}.

\bibitem{2022ApJ...934..184R}
{Ronchi}, M., {Rea}, N., {Graber}, V., \& {Hurley-Walker}, N.,
  \,\href{https://doi.org/10.3847/1538-4357/ac7cec}{\emph{\apj} {\bfseries 934}
  (2022) 184}.

\bibitem{2022MNRAS.509L..31P}
{Pelisoli}, I., {Marsh}, T.~R., {Dhillon}, V.~S., {et~al.},
  \,\href{https://doi.org/10.1093/mnrasl/slab116}{\emph{\mnras} {\bfseries 509}
  (2022) L31}.

\bibitem{2023MNRAS.523.3043R}
{Rigoselli}, M., {De Grandis}, D., {Mereghetti}, S., \& {Malacaria}, C.,
  \,\href{https://doi.org/10.1093/mnras/stad1611}{\emph{\mnras} {\bfseries 523}
  (2023) 3043}.

\bibitem{2021ApJ...923L...6K}
{Kilic}, M., {Kosakowski}, A., {Moss}, A.~G., {Bergeron}, P., \& {Conly},
  A.~A., \,\href{https://doi.org/10.3847/2041-8213/ac3b60}{\emph{\apjl}
  {\bfseries 923} (2021) L6}.

\bibitem{2001ApJ...556..987H}
{Harding}, A.~K. \& {Muslimov}, A.~G.,
  \,\href{https://doi.org/10.1086/321589}{\emph{\apj} {\bfseries 556} (2001)
  987}.

\bibitem{2002ApJ...568..862H}
{Harding}, A.~K. \& {Muslimov}, A.~G.,
  \,\href{https://doi.org/10.1086/338985}{\emph{\apj} {\bfseries 568} (2002)
  862}.

\bibitem{1986ApJ...300..500C}
{Cheng}, K.~S., {Ho}, C., \& {Ruderman}, M.,
  \,\href{https://doi.org/10.1086/163829}{\emph{\apj} {\bfseries 300} (1986)
  500}.

\bibitem{1986ApJ...300..522C}
{Cheng}, K.~S., {Ho}, C., \& {Ruderman}, M.,
  \,\href{https://doi.org/10.1086/163830}{\emph{\apj} {\bfseries 300} (1986)
  522}.

\bibitem{2018A&A...615A..73R}
{Rigoselli}, M. \& {Mereghetti}, S.,
  \,\href{https://doi.org/10.1051/0004-6361/201732408}{\emph{\aap} {\bfseries
  615} (2018) A73}.

\bibitem{2018ApJ...865..116I}
{Igoshev}, A.~P., {Tsygankov}, S.~S., {Rigoselli}, M., {et~al.},
  \,\href{https://doi.org/10.3847/1538-4357/aadd93}{\emph{\apj} {\bfseries 865}
  (2018) 116}.

\bibitem{2018ApJ...869...97A}
{Arumugasamy}, P., {Kargaltsev}, O., {Posselt}, B., {Pavlov}, G.~G., \& {Hare},
  J., \,\href{https://doi.org/10.3847/1538-4357/aaec69}{\emph{\apj} {\bfseries
  869} (2018) 97}.

\bibitem{2003A&A...407..315G}
{Gil}, J., {Melikidze}, G.~I., \& {Geppert}, U.,
  \,\href{https://doi.org/10.1051/0004-6361:20030854}{\emph{\aap} {\bfseries
  407} (2003) 315}.

\bibitem{2008ApJ...686..497G}
{Gil}, J., {Haberl}, F., {Melikidze}, G., {et~al.},
  \,\href{https://doi.org/10.1086/590657}{\emph{\apj} {\bfseries 686} (2008)
  497}.

\bibitem{2019ApJ...887L..26B}
{Bogdanov}, S., {Lamb}, F.~K., {Mahmoodifar}, S., {et~al.},
  \,\href{https://doi.org/10.3847/2041-8213/ab5968}{\emph{\apjl} {\bfseries
  887} (2019) L26}.

\bibitem{2019ApJ...887L..23B}
{Bilous}, A.~V., {Watts}, A.~L., {Harding}, A.~K., {et~al.},
  \,\href{https://doi.org/10.3847/2041-8213/ab53e7}{\emph{\apjl} {\bfseries
  887} (2019) L23}.

\bibitem{2019ApJ...887L..21R}
{Riley}, T.~E., {Watts}, A.~L., {Bogdanov}, S., {et~al.},
  \,\href{https://doi.org/10.3847/2041-8213/ab481c}{\emph{\apjl} {\bfseries
  887} (2019) L21}.

\bibitem{2023arXiv230903620P}
{P{\'e}tri}, J., {Guillot}, S., {Guillemot}, L., {et~al.},
  \,\href{https://doi.org/10.48550/arXiv.2309.03620}{\emph{arXiv e-prints}
  (2023) arXiv:2309.03620}.

\bibitem{zav04}
{Zavlin}, V.~E. \& {Pavlov}, G.~G.,
  \,\href{https://doi.org/10.1086/424894}{\emph{\apj} {\bfseries 616} (2004)
  452}.

\bibitem{her18}
{Hermsen}, W., {Kuiper}, L., {Basu}, R., {et~al.},
  \,\href{https://doi.org/10.1093/mnras/sty2075}{\emph{\mnras} {\bfseries 480}
  (2018) 3655}.

\bibitem{2019ApJ...872...15R}
{Rigoselli}, M., {Mereghetti}, S., {Turolla}, R., {et~al.},
  \,\href{https://doi.org/10.3847/1538-4357/aafac7}{\emph{\apj} {\bfseries 872}
  (2019) 15}.

\bibitem{2012Sci...337..946K}
{Kargaltsev}, O., {Durant}, M., {Misanovic}, Z., \& {Pavlov}, G.~G.,
  \,\href{https://doi.org/10.1126/science.1221378}{\emph{Science} {\bfseries
  337} (2012) 946}.

\bibitem{2022MNRAS.513.3113R}
{Rigoselli}, M., {Mereghetti}, S., {Anzuinelli}, S., {et~al.},
  \,\href{https://doi.org/10.1093/mnras/stac1130}{\emph{\mnras} {\bfseries 513}
  (2022) 3113}.

\bibitem{aru19}
{Arumugasamy}, P. \& {Mitra}, D.,
  \,\href{https://doi.org/10.1093/mnras/stz2299}{\emph{\mnras} {\bfseries 489}
  (2019) 4589}.

\bibitem{mis08}
{Misanovic}, Z., {Pavlov}, G.~G., \& {Garmire}, G.~P.,
  \,\href{https://doi.org/10.1086/590949}{\emph{\apj} {\bfseries 685} (2008)
  1129}.

\bibitem{2010ApJ...725L...6C}
{Caraveo}, P.~A., {De Luca}, A., {Marelli}, M., {et~al.},
  \,\href{https://doi.org/10.1088/2041-8205/725/1/L6}{\emph{\apjl} {\bfseries
  725} (2010) L6}.

\bibitem{2013ApJ...765...36M}
{Marelli}, M., {De Luca}, A., {Salvetti}, D., {et~al.},
  \,\href{https://doi.org/10.1088/0004-637X/765/1/36}{\emph{\apj} {\bfseries
  765} (2013) 36}.

\bibitem{2007ApJ...654..487N}
{Ng}, C.~Y., {Romani}, R.~W., {Brisken}, W.~F., {Chatterjee}, S., \& {Kramer},
  M., \,\href{https://doi.org/10.1086/510576}{\emph{\apj} {\bfseries 654}
  (2007) 487}.

\bibitem{2020MNRAS.493.1874D}
{Danilenko}, A., {Karpova}, A., {Ofengeim}, D., {Shibanov}, Y., \& {Zyuzin},
  D., \,\href{https://doi.org/10.1093/mnras/staa287}{\emph{\mnras} {\bfseries
  493} (2020) 1874}.

\bibitem{2014ApJ...793...88M}
{Mori}, K., {Gotthelf}, E.~V., {Dufour}, F., {et~al.},
  \,\href{https://doi.org/10.1088/0004-637X/793/2/88}{\emph{\apj} {\bfseries
  793} (2014) 88}.

\bibitem{2007ApJ...669..570M}
{Manzali}, A., {De Luca}, A., \& {Caraveo}, P.~A.,
  \,\href{https://doi.org/10.1086/521387}{\emph{\apj} {\bfseries 669} (2007)
  570}.

\bibitem{2012ApJ...744...81C}
{Chang}, C., {Pavlov}, G.~G., {Kargaltsev}, O., \& {Shibanov}, Y.~A.,
  \,\href{https://doi.org/10.1088/0004-637X/744/2/81}{\emph{\apj} {\bfseries
  744} (2012) 81}.

\bibitem{mcg04}
{McGowan}, K.~E., {Zane}, S., {Cropper}, M., {et~al.},
  \,\href{https://doi.org/10.1086/379787}{\emph{\apj} {\bfseries 600} (2004)
  343}.

\bibitem{2014ApJ...790...51M}
{Marelli}, M., {Belfiore}, A., {Saz Parkinson}, P., {et~al.},
  \,\href{https://doi.org/10.1088/0004-637X/790/1/51}{\emph{\apj} {\bfseries
  790} (2014) 51}.

\bibitem{her17}
{Hermsen}, W., {Kuiper}, L., {Hessels}, J.~W.~T., {et~al.},
  \,\href{https://doi.org/10.1093/mnras/stw3135}{\emph{\mnras} {\bfseries 466}
  (2017) 1688}.

\bibitem{2014ApJ...793L...8L}
{Lin}, L.~C.~C., {Hui}, C.~Y., {Li}, K.~T., {et~al.},
  \,\href{https://doi.org/10.1088/2041-8205/793/1/L8}{\emph{\apjl} {\bfseries
  793} (2014) L8}.

\bibitem{2021MNRAS.501.4998Z}
{Zyuzin}, D.~A., {Karpova}, A.~V., {Shibanov}, Y.~A., {Potekhin}, A.~Y., \&
  {Suleimanov}, V.~F.,
  \,\href{https://doi.org/10.1093/mnras/staa3991}{\emph{\mnras} {\bfseries 501}
  (2021) 4998}.

\bibitem{2012ApJ...761...65N}
{Ng}, C.~Y., {Kaspi}, V.~M., {Ho}, W.~C.~G., {et~al.},
  \,\href{https://doi.org/10.1088/0004-637X/761/1/65}{\emph{\apj} {\bfseries
  761} (2012) 65}.

\bibitem{mil13}
{Miller}, J.~J., {McLaughlin}, M.~A., {Rea}, N., {et~al.},
  \,\href{https://doi.org/10.1088/0004-637X/776/2/104}{\emph{\apj} {\bfseries
  776} (2013) 104}.

\bibitem{yon19}
{Yoneyama}, T., {Hayashida}, K., {Nakajima}, H., \& {Matsumoto}, H.,
  \,\href{https://doi.org/10.1093/pasj/psy135}{\emph{\pasj} {\bfseries 71}
  (2019) 17}.

\bibitem{2010ApJ...709..436H}
{Halpern}, J.~P. \& {Gotthelf}, E.~V.,
  \,\href{https://doi.org/10.1088/0004-637X/709/1/436}{\emph{\apj} {\bfseries
  709} (2010) 436}.

\bibitem{2004A&A...427..199C}
{Cassam-Chena{\"\i}}, G., {Decourchelle}, A., {Ballet}, J., {et~al.},
  \,\href{https://doi.org/10.1051/0004-6361:20041154}{\emph{\aap} {\bfseries
  427} (2004) 199}.

\bibitem{2011ApJ...731...70L}
{Lovchinsky}, I., {Slane}, P., {Gaensler}, B.~M., {et~al.},
  \,\href{https://doi.org/10.1088/0004-637X/731/1/70}{\emph{\apj} {\bfseries
  731} (2011) 70}.

\bibitem{2010ApJ...712..790T}
{Tian}, W.~W., {Li}, Z., {Leahy}, D.~A., {et~al.},
  \,\href{https://doi.org/10.1088/0004-637X/712/2/790}{\emph{\apj} {\bfseries
  712} (2010) 790}.

\bibitem{pav09}
{Pavlov}, G.~G. \& {Luna}, G.~J.~M.,
  \,\href{https://doi.org/10.1088/0004-637X/703/1/910}{\emph{\apj} {\bfseries
  703} (2009) 910}.

\bibitem{gre83}
{Greenstein}, G. \& {Hartke}, G.~J.,
  \,\href{https://doi.org/10.1086/161195}{\emph{\apj} {\bfseries 271} (1983)
  283}.

\bibitem{1985MNRAS.213..313H}
{Hernquist}, L., \,\href{https://doi.org/10.1093/mnras/213.2.313}{\emph{\mnras}
  {\bfseries 213} (1985) 313}.

\bibitem{tur13}
{Turolla}, R. \& {Nobili}, L.,
  \,\href{https://doi.org/10.1088/0004-637X/768/2/147}{\emph{\apj} {\bfseries
  768} (2013) 147}.

\bibitem{per06a}
{P{\'e}rez-Azor{\'{\i}}n}, J.~F., {Miralles}, J.~A., \& {Pons}, J.~A.,
  \,\href{https://doi.org/10.1051/0004-6361:20054403}{\emph{\aap} {\bfseries
  451} (2006) 1009}.

\bibitem{per06b}
{P{\'e}rez-Azor{\'\i}n}, J.~F., {Pons}, J.~A., {Miralles}, J.~A., \&
  {Miniutti}, G.,
  \,\href{https://doi.org/10.1051/0004-6361:20065827}{\emph{\aap} {\bfseries
  459} (2006) 175}.

\bibitem{gep06}
{Geppert}, U., {K{\"u}ker}, M., \& {Page}, D.,
  \,\href{https://doi.org/10.1051/0004-6361:20054696}{\emph{\aap} {\bfseries
  457} (2006) 937}.

\bibitem{2019LRCA....5....3P}
{Pons}, J.~A. \& {Vigan{\`o}}, D.,
  \,\href{https://doi.org/10.1007/s41115-019-0006-7}{\emph{LRCA} {\bfseries 5}
  (2019) 3}.

\bibitem{2021MNRAS.506.4593Y}
{Yakovlev}, D.~G.,
  \,\href{https://doi.org/10.1093/mnras/stab2077}{\emph{\mnras} {\bfseries 506}
  (2021) 4593}.

\bibitem{2020ApJ...903...40D}
{De Grandis}, D., {Turolla}, R., {Wood}, T.~S., {et~al.},
  \,\href{https://doi.org/10.3847/1538-4357/abb6f9}{\emph{\apj} {\bfseries 903}
  (2020) 40}.

\bibitem{2021ApJ...914..118D}
{De Grandis}, D., {Taverna}, R., {Turolla}, R., {et~al.},
  \,\href{https://doi.org/10.3847/1538-4357/abfdac}{\emph{\apj} {\bfseries 914}
  (2021) 118}.

\bibitem{2021NatAs...5..145I}
{Igoshev}, A.~P., {Hollerbach}, R., {Wood}, T., \& {Gourgouliatos}, K.~N.,
  \,\href{https://doi.org/10.1038/s41550-020-01220-z}{\emph{Nature Astronomy}
  {\bfseries 5} (2021) 145}.

\end{thebibliography}\endgroup





\end{document}